\documentclass[11pt,a4paper]{article}
\usepackage{amssymb}
\usepackage{amsmath}
\usepackage{color}
\usepackage[colorlinks,linkcolor=blue,citecolor=red]{hyperref}

\newcommand{\R}{\mathbb{R}}

\newcommand{\bbbr}{\mathbb R}

\newtheorem{theorem}{Theorem}
\newtheorem{theorema}{Theorem}
\newtheorem{prop}[theorema]{Proposition}

\newcommand{\comm}[1]{}

\renewcommand\a{\alpha}

\newcommand\E{{\mathcal E}}

\newcommand\La{\Lambda}

\newcommand\op[1]{\mathop{\rm #1}\nolimits}
\newcommand\ot{\otimes}
\newcommand\p{\partial}

\newcommand\po{$\!\!\!{\text{\bf.}}$ }

\begin{document}

\title{Dispersionless integrable hierarchies and $GL(2, \bbbr)$ geometry}

\author{E.V. Ferapontov, B. Kruglikov}
     \date{}
     \maketitle
     \vspace{-5mm}
\begin{center}
{\small
Department of Mathematical Sciences \\ Loughborough University \\
Loughborough, Leicestershire LE11 3TU \\ United Kingdom, \\
  \ \\  \ \\
Institute of Mathematics and Statistics\\
Faculty of Science\\
University of Troms\o\\
Troms\o\ 90-37, Norway\\
[2ex]
e-mails: \\[1ex]  \texttt{E.V.Ferapontov@lboro.ac.uk}\\
\texttt{Boris.Kruglikov@uit.no} }

\bigskip

\end{center}

\begin{abstract}

Paraconformal or $GL(2,\bbbr)$ geometry on an $n$-dimensional  manifold $M$ is defined by a field of
rational normal curves of degree $n-1$ in the projectivised cotangent bundle $\mathbb{P} T^*M$. Such geometry
is known to arise on solution spaces of ODEs with vanishing W\"unschmann (Doubrov-Wilczynski) invariants.
In this paper we discuss yet another natural source of $GL(2,\bbbr)$ structures, namely dispersionless
integrable hierarchies of PDEs such as the dispersionless Kadomtsev-Petviashvili (dKP) hierarchy. In the latter
context, $GL(2,\bbbr)$ structures  coincide with the characteristic variety (principal symbol) of the hierarchy.

Dispersionless hierarchies provide explicit examples of  particularly interesting classes of involutive
$GL(2,\bbbr)$ structures studied in the literature. Thus, we obtain torsion-free $GL(2, \bbbr)$ structures
of  Bryant \cite{Bryant} that appeared in the context of exotic holonomy  in dimension four,
as well as totally geodesic  $GL(2,\bbbr)$ structures of  Krynski \cite{Krynski}.
The latter possess  a compatible  affine connection (with torsion)
and a two-parameter family of  totally geodesic $\alpha$-manifolds (coming from the dispersionless Lax equations),
which makes them a natural generalisation of the Einstein-Weyl geometry.

Our main result states that involutive $GL(2,\bbbr)$ structures are governed by a dispersionless integrable system
whose general local solution depends on $2n-4$ arbitrary functions of 3 variables.
This establishes integrability of the system of W\"unschmann conditions.

\medskip


\end{abstract}

\newpage


\tableofcontents

\section{Introduction}

\subsection{$GL(2, \bbbr)$ geometry}

On an $n$-dimensional manifold $M$, a $GL(2, \bbbr)$ geometry (also known as paraconformal geometry \cite{DT},  a rational normal structure \cite{Bryant2}, or a special case of the cone structure \cite{Gin})
is defined by a field of rational normal curves of degree $n-1$
in the projectivised cotangent bundle $\mathbb{P} T^*M$.  Equivalently, it can be viewed as a field of 1-forms $\omega(\lambda)$  polynomial of degree $n-1$ in $\lambda$,
\begin{equation}
\omega (\lambda)=\omega_0+\lambda \omega_1+ \dots + \lambda^{n-1}\omega_{n-1},
\label{o}
\end{equation}
where $\omega_i$ is a basis of 1-forms  (a coframe) on $M$. The parameter $\lambda$ and the 1-form $\omega(\lambda)$ are defined up to transformations  $\lambda \to  \frac{a\lambda+b}{c\lambda +d}, \ \omega(\lambda) \to r(c\lambda+d)^{n-1}\omega(\lambda)$, where $a, b, c, d, r$  are arbitrary smooth functions on $M$
such that $ad-bc\ne 0, \ r\ne 0$. Without any loss of generality we can assume $ad-bc=1$.

Conventionally, a $GL(2, \bbbr)$ geometry is defined by a field of rational normal curves  in the projectivised {\it tangent} bundle $\mathbb{P} TM$. Our choice of the cotangent bundle is motivated by the fact that characteristic varieties of PDEs, which will be our main source of $GL(2, \bbbr)$ structures, are subvarieties of $\mathbb{P} T^*M$. In any case, both pictures are  projectively dual: the equation $\omega(\lambda)=0$ defines a one-parameter family of hyperplanes that osculate a dual rational normal curve $\tilde \omega(\lambda)\subset \mathbb{P} TM$. Below we discuss some of the most natural occurrences of $GL(2, \bbbr)$ structures.

\smallskip

\noindent {\bf Poisson geometry}: Given a generic pair of compatible Poisson bivectors $\eta_1, \eta_2$ of Kronecker type on an odd-dimensional manifold $N^{2k+1}$, there is a canonical $GL(2, \bbbr)$ structure on the base $M^{k+1}$ (leaf space) of the corresponding action foliation (see \cite{Zakharevich}).
 As shown by Gelfand and Zakharevich such structures, also known as Veronese webs,
 arise in the theory of bi-Hamiltonian integrable systems  \cite{GZ}.

\smallskip

\noindent {\bf Exotic holonomy}: It was observed by Bryant in \cite{Bryant} that, in four dimensions, there exist torsion-free affine connections whose holonomy group is  the irreducible representation of  $GL(2, \bbbr)$. Such connections give rise to canonically defined parallel $GL(2, \bbbr)$ structures. Historically, this was the first example of an `exotic' holonomy not appearing on the Berger  list \cite{Berger}, we refer to \cite{Bryant1, Merkulov} for the development of the holonomy problem.

\smallskip

\noindent {\bf Submanifolds in Grassmannians}: Let $M$ be a submanifold of the Grassmannian $Gr(k, n)$. The flat Segre structure of $Gr(k, n)$ induces on $M$ a generalised conformal structure. Particular instances of this construction result in a $GL(2, \bbbr)$ geometry on $M$.

Thus, let $M^4$ be a fourfold in the Grassmannian ${ Gr}(3, 5)$. The  flat Segre structure of $Gr(3, 5)$ induces a field of twisted cubics on $\mathbb{P} TM^4$, that is, a $GL(2, \bbbr)$ structure on $M^4$. These structures were  investigated in \cite{DFKN} in the context of integrability in Grassmann geometries.

Similarly, let $\Lambda(3)$ be the  Grassmannian of 3-dimensional Lagrangian subspaces of a 6-dimensional symplectic space. Given a hypersurface   $M^5\subset \Lambda(3)$, the  flat Veronese structure of  $\Lambda(3)$ induces a $GL(2, \bbbr)$ structure on $M^5$. Such structures were discussed in \cite{FHK, AS} in the context of integrability of dispersionless Hirota type equations.

\smallskip

\noindent {\bf Algebraic geometry}: Given a compact complex surface $X$  and a rational curve $C\subset X$  with the normal bundle $\nu \simeq{\cal O}(n)$, the results of Kodaira \cite{Kodaira} show that there is a complex-analytic $(n+1)$-dimensional moduli space $M$ consisting of deformations of $C$, which carries a canonical $GL(2, \bbbr)$ structure. This was studied in detail by Hitchin \cite{Hitchin} for $n=2$
(using \'E.~Cartan's work on Einstein-Weyl geometry) and by Bryant \cite{Bryant} for $n=3$. The case of general $n$ was discussed by Dunajski, Tod \cite{DT} and Krynski \cite{Krynski}.
The construction generalises to the case when $X$ is a holomorphic contact 3-fold and $C \subset X$ is a contact rational curve with the normal bundle  $\nu \simeq{\cal O}(n-1)\oplus{\cal O}(n-1)$ \cite{Bryant, DT, Bryant2}.

\smallskip

\noindent {\bf Ordinary differential equations}: For every scalar (higher order) ODE with  vanishing W\"un\-schmann (Doubrov-Wilczynski) invariants, the space $M$ of its solutions is canonically endowed with a $GL(2, \bbbr)$ structure. ODEs of this type have been thoroughly investigated in the literature, see e.g. \cite{DT, Doubrov, Nurowski, God, DK, Krynski} and references therein.

\smallskip

\noindent {\bf Dispersionless integrable hierarchies}: Given a dispersionless integrable hierarchy of PDEs, it will be demonstrated in this paper that the corresponding characteristic variety (zero locus of the principal symbol) determines canonically a $GL(2, \bbbr)$ structure on every solution. In a somewhat different language examples of this type  appeared in \cite{D, Krynski}, although  the observation that these structures coincide with the characteristic variety is apparently new. We will show that the $GL(2, \bbbr)$ structures  appearing on solutions to integrable hierarchies are not arbitrary, and must satisfy an important property of {\it involutivity}.

\subsection{Involutive $GL(2,\bbbr)$ structures and dispersionless hierarchies}
\label{sec:invol}

For every $x\in M$, the equation $\omega(\lambda)=0$ defines a 1-parameter family of hyperplanes in $T_xM$
parametrised by $\lambda$; these are known as $\alpha$-hyperplanes.   A codimension one submanifold of $M$ is said
to be an {\it $\alpha$-manifold} if all its tangent spaces are $\alpha$-hyperplanes \cite{Krynski}.

\medskip

\noindent {\bf Definition.}
A $GL(2, \bbbr)$ structure is said to be {\em involutive\/} \cite{Bryant2} or {\it $\alpha$-integrable} \cite{Krynski}
if every $\alpha$-hyperplane is tangential to some $\alpha$-manifold.

\medskip

We will relate different approaches to involutivity in Section \ref{sec:ext}.
One can show that $\alpha$-manifolds of an involutive $GL(2, \bbbr)$ structure depend on 1 arbitrary function of 1
variable; see Section \ref{sec:alpha-num}. The existence of  $\alpha$-manifolds suggests that involutive
$GL(2, \bbbr)$ structures are amenable to twistor-theoretic methods, cf. \cite{Gin}.

In particular, $GL(2, \bbbr)$ structures that arise on solution spaces of ODEs with vanishing W\"unschmann invariants are
involutive. It was shown in \cite{Krynski} that conversely, every involutive ($\alpha$-integrable) $GL(2, \bbbr)$ structure
can be obtained from an ODE of this type.  Four-dimensional involutive  $GL(2, \bbbr)$ structures  were extensively studied
in \cite{Bryant} in the context of  exotic holonomy. These investigations were developed further in
\cite{DT, Doubrov, Nurowski, God, DK}.



Our main observation is that involutive $GL(2, \bbbr)$ structures are induced,  as  characteristic varieties, on solutions to 
dispersionless integrable hierarchies of PDEs.
Moreover, $\alpha$-manifolds can be obtained as projections of integral
manifolds of the associated  dispersionless Lax equations.

The following example is based on \cite{Zakharevich, DK, Krynski}.  Equations of the Veronese web hierarchy have the form
 \begin{equation}
\begin{array}{c}
(c_i-c_j) u_ku_{ij}+(c_j-c_k) u_iu_{jk}+(c_k-c_i) u_j u_{ik}=0,
\end{array}
\label{Ver}
 \end{equation}
one equation for every triple $(i, j, k)$ of distinct indices. Here $u$ is a function on the $n$-dimensional manifold $M$
with local coordinates $x^1, \dots, x^{n}$, coefficients $c_i$ are pairwise distinct constants, and $u_i=u_{x^i}$ denote
partial derivatives.

The term  `hierarchy' refers to the fact that the overdetermined system (\ref{Ver}) is in involution
for every $n$ so that any two equations can be viewed as Lie-B\"acklund symmetries of each other:
if we take two different triples and unite the indices, then the system of equations of type \eqref{Ver}
corresponding to all sub-triples of the union is compatible.

The characteristic variety  of system (\ref{Ver}) is defined by a system of quadrics,
$$
\begin{array}{c}
(c_i-c_j) u_kp_ip_j+(c_j-c_k) u_ip_jp_k+(c_k-c_i) u_j p_ip_k=0,
\end{array}
$$
which specify a rational normal curve in $\mathbb{P}T^*M$  parametrised as $p_i=\frac{u_i}{\lambda-c_i}$ (the ideal of a rational normal curve is generated by quadrics, see e.g. \cite{Harris}).
Explicitly,
\begin{equation}
\omega(\lambda)=p_idx^i= \sum \frac{u_i}{\lambda-c_i} \ dx^i;
\label{Vero}
\end{equation}
note that  expression (\ref{Vero}) takes form (\ref{o}) on clearing the denominators
(since only the conformal
class of $\omega(\lambda)$ is essential we will not make a distinction in what follows).
This supplies $M$ with a $GL(2, \bbbr)$ geometry which depends on the solution $u$
(otherwise said: $GL(2, \bbbr)$ geometry on the solution $u$ considered as a submanifold
$\op{graph}(u)\subset M\times\R$).

System (\ref{Ver}) is equivalent to the commutativity conditions of the following vector fields ($\lambda$=const),
$$
\partial_{x^j}-\frac{\lambda-c_1}{\lambda-c_j}\frac{u_j}{u_1}\partial_{x^1}, ~~~ 1<j\leq n.
$$
Such $\lambda$-dependent vector fields are said to define a dispersionless Lax representation for system (\ref{Ver}).
Note that these vector fields are annihilated by $\omega(\lambda)$. Their integral manifolds  supply $M$ with a two-parameter family of $\alpha$-manifolds.
Thus,  $GL(2, \bbbr)$  structure (\ref{Vero}) is involutive.
 Equivalently, the commutativity of these vector fields  can be interpreted as the involutivity of the corresponding corank 2 vector distribution on the $(n+1)$-dimensional manifold $\hat M$ with coordinates $x^1, \dots, x^{n}, \lambda$, known as the correspondence space. The (complexified)  space of integral manifolds of this distribution  plays  important role in the twistor-theoretic approach to the Veronese web hierarchy.

In Section \ref{sec:ex} we provide further examples of involutive $GL(2, \bbbr)$ structures supported on solutions to other well-known dispersionless integrable hierarchies.

\subsection{Affine connections associated with involutive $GL(2, \bbbr)$ structures}
\label{sec:con}

There are several types of canonical connections defined on the tangent bundle of a manifold $M$ that can be naturally associated with a
$GL(2, \bbbr)$ structure on  $M$.
Recall that an affine connection $\nabla$ is said to be compatible with a $GL(2, \bbbr)$ structure (paraconformal or $GL(2, \bbbr)$ connection),  if  for every ${\rm v}\!\in TM$
 \begin{equation}
\nabla_{\rm\!v\,}\omega(\lambda)\in {\rm span}\langle\omega(\lambda), \omega'(\lambda)\rangle,
\label{K}
 \end{equation}
where prime denotes differentiation by $\lambda$, see \cite{Krynski}. Condition (\ref{K}) means that the parallel transport defined by $\nabla$ preserves rational
normal cones of the $GL(2, \bbbr)$ structure.
Equivalently, identifying quadratic equations from the ideal of the rational normal curve
$\omega(\lambda)$ with symmetric bivectors $g_{s}$ on $M$ and denoting
$g= {\rm span} \langle g_s\rangle$, we can represent (\ref{K}) as
$\nabla_{\rm\!v\,} g= g$ $\forall {\rm\,v}\!\neq0$.

Condition (\ref{K}) alone does not specify $\nabla$ uniquely: for this, additional constraints should be imposed. In what follows we discuss four types of canonical connections associated with involutive $GL(2, \bbbr)$ structures,
of which the first two are based on the previous works and  do not exist universally, while the
other two are new and  exist for all dispersionless integrable hierarchies studied so far
(let us stress that there exist no general theory or complete description of such hierarchies).
We use the convention $\nabla_j\partial_k=\Gamma^i_{jk}\partial_i$.

\subsubsection*{Torsion-free  $GL(2, \bbbr)$ connection in 4D}

Torsion-free $GL(2, \bbbr)$ connections can only exist  in four dimensions. Indeed, based on the Berger criteria, it was shown  in \cite{Bryant} that there exist no non-trivial torsion-free  $GL(2, \bbbr)$ connections in higher dimensions.  On the contrary, in four dimensions,  involutivity of a $GL(2, \bbbr)$ structure is equivalent to the existence of  a  torsion-free $GL(2, \bbbr)$ connection.

Since  $GL(2, \bbbr)$ structures coming from principal symbols of  dispersionless integrable hierarchies are  automatically involutive (due to the existence of a Lax representation), we obtain an abundance of explicit examples of torsion-free  $GL(2, \bbbr)$ connections in four dimensions parametrised by solutions to some well-known integrable PDEs, see Section \ref{sec:ex}.

For the Veronese web hierarchy, the Christoffel symbols of the torsion-free $GL(2, \bbbr)$ connection associated with four-dimensional $GL(2, \bbbr)$ structure (\ref{Vero}) are computed to be equal to
$$
\Gamma^i_{ii}=\frac{u_{ii}}{u_i}-\frac{1}{9}\sum_{j\ne i}\frac{(c_{ik}c_{jl}+c_{il}c_{jk})^2}{c_{ik}c_{il}c_{jk}c_{jl}}\frac{u_{ij}}{u_j}, ~~~~~ \Gamma^j_{ii}=\frac{1}{9}\frac{c_{jk}c_{jl}}{c_{ij}c_{lk}}\frac{u_i}{u_j} \left(\frac{u_{ik}}{u_k}-\frac{u_{il}}{u_l}\right),
$$
$$
\Gamma^i_{ij}=\frac{1}{3}\frac{u_{ij}}{u_i} -\frac{1}{9} \left(1+\frac{c_{ik}c_{jl}}{c_{ij}c_{kl}}\right)\frac{u_{jl}}{u_l}-\frac{1}{9} \left(1+\frac{c_{il}c_{jk}}{c_{ij}c_{lk}}\right)\frac{u_{jk}}{u_k}, ~~~~~
\Gamma^j_{ik}=\frac{1}{9}\frac{c_{lj}}{c_{lk}}\frac{u_k}{u_j}\left(\frac{u_{ik}}{u_k}-\frac{u_{il}}{u_l}\right),
$$
here $c_{ij}=c_i-c_j$, and $i, j, k, l$ are pairwise distinct indices taking values $1, \dots, 4$.

\subsubsection*{Totally geodesic $GL(2, \bbbr)$ connections}

A particularly interesting subclass of  involutive $GL(2, \bbbr)$ structures was  introduced by Krynski in \cite{Krynski}: such structures
possess  a $GL(2, \bbbr)$ connection (with torsion) and a two-parameter family of  totally geodesic $\alpha$-manifolds.
We will refer to such structures/connections  as  {\it totally geodesic}  $GL(2, \bbbr)$ structures/connections, respectively.

The requirement that  $\nabla$ is a  totally geodesic $GL(2, \bbbr)$ connection specifies it
up to transformation $\Gamma^i_{jk}\to \Gamma^i_{jk}+\phi_j\delta^i_k$
for a covector $\phi$. This freedom can be eliminated by requiring that the torsion $T_\nabla$
is trace-free, $T^i_{ik}=0$. In what follows this will be included into the totally geodesic condition.
For  $GL(2, \bbbr)$ structures (\ref{Vero}) coming from the Veronese web hierarchy, the condition
$\op{tr}\bigl(T_\nabla(\cdot,X)\bigr)=0$ is equivalent to the constraint  $T_\nabla(\tilde \omega(\lambda), \tilde\omega'(\lambda))\in {\rm span} \langle\tilde \omega(\lambda)\rangle$ used in \cite{Krynski}.


Examples of totally geodesic $GL(2, \bbbr)$ structures include the following:
\begin{itemize}

\item {\it Four-dimensional}
$GL(2, \bbbr)$ structures arising, as characteristic varieties,
on solutions to various integrable hierarchies (see Appendix \ref{App.B}). We emphasise that,
in general, this is a merely 4-dimensional phenomenon. For instance, 5-dimensional $GL(2, \bbbr)$
structures associated with the dKP hierarchy do not possess totally geodesic $GL(2, \bbbr)$ connections.

\item {\it Multi-dimensional }
$GL(2, \bbbr)$ structures arising, as characteristic varieties, on solutions to {\it linearly degenerate}
integrable hierarchies (those having no $\partial_\lambda$ in the Lax fields,
such as the Veronese web hierarchy and the `universal' hierarchy). The two-parameter family of
totally geodesic $\alpha$-manifolds is the projection of integral manifolds of the Lax distribution.

\end{itemize}

It was shown in \cite{Krynski} that totally geodesic $GL(2, \bbbr)$ connections $\nabla$ satisfy the following multi-dimen\-si\-onal
{\it generalized Einstein-Weyl property\/}. Namely, the symmetrised Ricci tensor of such $\nabla$
belongs to the span $\tilde g$ of symmetric bivectors defining the dual rational normal curve $\tilde\omega(\lambda)$:
$Ric_\nabla^\text{\rm sym}\in\tilde{g}$. Note that in 3D this is precisely the classical Einstein-Weyl condition.

\subsubsection*{Normal $GL(2, \bbbr)$ connections}


We call a $GL(2, \bbbr)$ connection $\nabla$ {\it normal} if its torsion satisfies the following properties:
 \begin{itemize}
\item[(i)] $T_\nabla$ is trace-free: ${\rm{tr}\bigl(T_\nabla(\cdot,X)\bigr)}=0$ \  $\forall X$;
\item[(ii)] $T_\nabla$ preserves $\alpha$-hyperplanes as a (2,1)-map:
$X,Y\in\omega(\lambda)^\perp$ $\Rightarrow$
$T_\nabla(X,Y)\in\omega(\lambda)^\perp$.
 \end{itemize}

Every totally geodesic $GL(2, \bbbr)$ connection is necessarily normal, although the converse is not true in general.
It turns out that for all hierarchies we investigated, the normal $GL(2, \bbbr)$ connection exists, and is unique (we point out that there are no totally geodesic connections associated with higher-dimensional  $GL(2, \bbbr)$ structures coming from the dKP and the Adler-Shabat hierarchies, starting from dimension 5).
The importance of normal $GL(2, \bbbr)$ connections lies in the fact that every such $\nabla$ satisfies the generalized Einstein-Weyl property.



The totally geodesic (and thus normal) $GL(2, \bbbr)$ connection associated with $GL(2, \bbbr)$ structure
(\ref{Vero}) of the Veronese web hierarchy is given by the formula
 $$
\nabla_j\partial_k=\left(\frac{u_{jk}}{u_k} +\phi_j \right)\partial_k, ~~~~ {\rm or} ~~~~ \Gamma^i_{jk}=\left(\frac{u_{jk}}{u_k} +\phi_j \right)\delta^i_k;
 $$
here the covector $\phi_j$ is still arbitrary  \cite{Krynski}.
It can be fixed uniquely by requiring the torsion to be trace-free:
 $$
\phi_j=-\frac{1}{n-1} \sum_{k\ne j}\frac{u_{jk}}{u_k}.
 $$

\subsubsection*{A canonical projective connection}

There exists yet another class of affine connections associated with involutive $GL(2, \bbbr)$ structures, namely, torsion-free connections possessing a two-parameter family of totally geodesic $\alpha$-mani\-folds;
note that they do not preserve the $GL(2, \bbbr)$ structure in general.

For $GL(2, \bbbr)$ structures defined by the characteristic varieties of dispersionless hierarchies,
the two-parameter family of totally geodesic $\alpha$-manifolds come from projections of integral manifolds
of the corresponding dispersionless Lax  equations.

The requirement that  $\nabla$ is a torsion-free  connection with a two-parameter family of totally geodesic $\alpha$-manifolds
specifies it uniquely up to projective equivalence,
$\Gamma^i_{jk}\to \Gamma^i_{jk}+\phi_j\delta^i_k+\phi_k\delta^i_j$, for a 1-form $\phi$.
Thus, we obtain a canonically defined  totally geodesic {\it projective} connection.

For the involutive $GL(2, \bbbr)$ structure (\ref{Vero}) of the Veronese web hierarchy, an affine representative
of this projective connection is computed to be equal to
$$
\nabla_j \partial_k=\frac{u_{jk}}{2}\left(\frac{\partial_j}{u_j}+\frac{\partial_k}{u_k}\right).
$$
On every solution, geodesics of this projective connection  (considered as unparametrized curves)
can be obtained by intersecting $n-2$ generic totally geodesic $\alpha$-manifolds.

\subsection{Summary of the main results}

In  Section \ref{sec:ex} we provide further explicit examples of involutive $GL(2, \bbbr)$ structures given by  characteristic varieties of various dispersionless integrable hierarchies, namely the dKP hierarchy, the `universal' hierarchy  of Martinez-Alonso and Shabat, and the consistent Adler-Shabat triples. In each case we calculate  Christoffel's symbols of the canonical connections discussed in Section \ref{sec:con} (these results are relegated to Appendix \ref{App.B}).

Section \ref{sec:gen} contains the main results of the paper. In Theorem \ref{Prop1} we demonstrate that
the general involutive  $GL(2, \bbbr)$ structure can be brought to the normal form
 \begin{equation}
\omega(\lambda)=\sum_{i=1}^n \frac{u_i}{\lambda-\frac{u_i}{v_i}}dx^i,
\label{form}
 \end{equation}
which can be reduced to (\ref{o}) by clearing denominators. Here   the functions $u$ and $v$  satisfy a system of  second-order PDEs, 2 equations for each quadruple of indices $1\leq i<j<k<l\leq n$:
 \begin{equation}
\mathop{\mathfrak{S}}\limits_{(jkl)} (a_i-a_j)(a_k-a_l)\left( \frac{2u_{ij}-(a_i+a_j)v_{ij}}{u_iu_j}+ \frac{2u_{kl}-(a_k+a_l)v_{kl}}{u_ku_l} \right)=0,
\label{un}
 \end{equation}
 \begin{equation}
\mathop{\mathfrak{S}}\limits_{(jkl)}(b_i-b_j)(b_k-b_l)\left( \frac{2v_{ij}-(b_i+b_j)u_{ij}}{v_iv_j}+ \frac{2v_{kl}-(b_k+b_l)u_{kl}}{v_kv_l} \right)=0,
\label{vn}
 \end{equation}
where $a_i=\frac{u_i}{v_i}, \ b_i=\frac{v_i}{u_i}$, and $\mathop{\mathfrak{S}}$ denotes cyclic summation
over the indicated indices.

In Theorem \ref{T2} we prove that  overdetermined system (\ref{un}), (\ref{vn}) is in involution, and its
characteristic variety is the tangential variety of the rational normal curve  $\omega(\lambda)$ given by (\ref{form}).
Since the degree of the tangential variety equals $2n-4$, we conclude that general involutive $GL(2, \bbbr)$ structures  depend (modulo diffeomorphisms) on $2n-4$ arbitrary functions of 3 variables.
For $n=4$ this reproduces the count in \cite{Bryant}; we also refer to \cite{KM} for an alternative PDE system
governing involutive $GL(2, \bbbr)$ structures for $n=4$.
For general $n$, the functional freedom of $2n-4$ arbitrary functions of 3 variables was announced by Robert Bryant
in a series of talks in the early 2000s \cite{Bryant2} (we thank him for sending us the slides), but no proofs
have appeared. Our proof is based on the formal theory of PDEs developed in recent years.

Finally, in Theorem \ref{T1} we show that equations (\ref{un}),\,(\ref{vn})
governing general involutive $GL(2, \bbbr)$ structures constitute a dispersionless integrable hierarchy
with Lax representation in parameter-dependent vector fields.


It was shown in \cite{Krynski}  that involutive $GL(2, \bbbr)$ structures are in one-to-one
correspondence with ODEs having vanishing W\"unschmann invariants.
Thus, integrability of system (\ref{un}),\,(\ref{vn}) implies integrability of the W\"unschmann conditions.

Our considerations are local. All results on the functional freedom in the general solution referring to the Cartan-K\"ahler theorem  hold in the analytic or formal categories.

\section{Examples of involutive $GL(2, \bbbr)$ structures}
\label{sec:ex}

In this section we give further  examples of involutive  $GL(2, \bbbr)$ structures arising on solutions of various dispersionless integrable hierarchies. Our main observation is that $GL(2, \bbbr)$ structures discussed in a similar context by Dunajski and Krynski in \cite{D, Krynski} are nothing but characteristic varieties of the corresponding PDEs. This makes the construction entirely explicit and intrinsic.

We mainly focus on $GL(2, \bbbr)$ geometry in four dimensions, defined by the first three equations
of the corresponding hierarchies. Higher-dimensional generalisations are then obtained by adding higher flows
(with higher time variables).  Christoffel's symbols of the  canonical connections associated with these
examples are presented in Appendix \ref{App.B}.

\subsection{$GL(2, \bbbr)$ structures via dKP hierarchy}

The first three equations of the dKP hierarchy have the form
 \begin{equation}
\begin{array}{c}
u_{xt}-u_{yy}-u_{x}u_{xx}=0, \\
u_{xz}-u_{yt}-u_{x}u_{xy}-u_{y}u_{xx}=0, \\
u_{yz}-u_{tt}+u_{x}^2u_{xx}-u_{y}u_{xy}=0.
\end{array}\label{dKP2}
 \end{equation}
Here $u$ is a function on the 4-dimensional manifold $M$ with local coordinates $x, y, t, z$.
The characteristic variety of this system is the intersection of three quadrics,
\begin{equation}
\begin{array}{c}
p_xp_t-p_y^2-u_{x}p_x^2=0,\\
p_xp_z-p_yp_t-u_{x}p_xp_y-u_{y}p_x^2=0,\\
 p_yp_z-p_t^2+u_{x}^2p_x^2-u_{y}p_xp_y=0,
 \end{array}
\label{dKPchar}
\end{equation}
which specify a rational normal curve (twisted cubic) in  $\mathbb{P}T^*M$ parametrised as
$$p_x=1,\ p_y=\lambda,\ p_t=\lambda^2+u_x, \ p_z=\lambda^3+2u_x\lambda+u_y,$$ so that
$$
\omega(\lambda)= dx+\lambda dy+(\lambda^2+u_x)dt+(\lambda^3+2u_x\lambda+u_y)dz.
$$
This supplies $M$ with a $GL(2, \bbbr)$ geometry which depends on the solution $u$. The occurrence of a rational normal curve in the theory of dKP hierarchy was also noted in \cite{Kon} in the context of coisotropic deformations of algebraic curves.
Equations (\ref{dKP2})   are equivalent to the commutativity conditions of the following vector fields,
\begin{equation}
\begin{array}{c}
\partial_y-\lambda \partial_x+u_{xx}\partial_{\lambda}, \\
\partial_t-(\lambda^2 +u_x)\partial_x+(\lambda u_{xx}+u_{xy})\partial_{\lambda}, \\
\partial_z-(\lambda^3+2u_x\lambda+u_y)\partial_x+(\lambda^2u_{xx}+\lambda u_{xy}+u_{xt}+u_xu_{xx})\partial_{\lambda},
\end{array}
\label{dKPLax}
\end{equation}
which constitute a dispersionless Lax representation. These vector fields live in the extended 5-dimensional space $\hat M$ with coordinates $x, y, t, z, \lambda$; note the explicit presence of  $\partial_{\lambda}$. Projecting integral manifolds of these vector fields from $\hat M$ to $M$ we obtain a two-parameter family of $\alpha$-manifolds of the corresponding $GL(2, \bbbr)$ structure, thus establishing its involutivity.

\medskip

Higher-dimensional generalisation of this construction can be obtained by taking higher flows of the dKP hierarchy,
$$
u_{i, j+1}-u_{j, i+1}+\sum_{k=1}^{i}u_{i-k}u_{jk}-\sum_{k=1}^ju_{j-k}u_{ik}=0, ~~~ 1\leq i< j,
$$
see e.g. \cite{KonMag}. For $(i, j)=(1, 2), (1, 3)$ and $(2, 3)$ this reproduces equations (\ref{dKP2}). Here we use the notation $u=u(x^1, x^2, x^3, x^4, \dots)$ where $x^1=x,\ x^2=y,\ x^3=t, \ x^4=z$, etc,  and subscripts of $u$ denote partial derivatives. The corresponding characteristic variety is the intersection of quadrics,
$$
p_ip_{j+1}-p_jp_{i+1}+\sum_{k=1}^{i}u_{i-k}p_jp_k-\sum_{k=1}^ju_{j-k}p_ip_k=0.
$$
It defines a rational normal curve; setting $p_1=1$ we can parametrise it  recurrently as
$$
p_{i+1}=\lambda p_i+\sum_{k=1}^{i-1}u_{i-k}p_k, ~~~ i\geq 1.
$$
Explicitly, this gives
$$
p_1=1, ~~~ p_2=\lambda, ~~~ p_3=\lambda^2+u_1, ~~~ p_4=\lambda^3+2u_1\lambda+u_2, ~~~ p_{5}=\lambda^4+3u_1\lambda^2+2u_2\lambda+u_3+u_1^2,
$$
etc. The dispersionless Lax representation of the dKP hierarchy is given by a family of involutive parameter-dependent vector fields
$$
X_i=\partial_{x^{i+1}}-\lambda \partial_{x^i}-\sum_{k=1}^{i-1}u_{i-k}\partial_{x^k}+u_{1i}\partial_{\lambda}, ~~~ i\geq 1.
$$

\subsection{$GL(2, \bbbr)$ structures via the universal hierarchy}

The first three equations of the universal hierarchy of Martinez-Alonso and Shabat \cite{Shabat} have the form
 \begin{equation}
\begin{array}{c}
u_{xt}-u_{yy}+u_{y}u_{xx}-u_xu_{xy}=0, \\
u_{xz}-u_{yt}+u_{t}u_{xx}-u_{x}u_{xt}=0, \\
u_{yz}-u_{tt}+u_{t}u_{xy}-u_{y}u_{xt}=0.
\end{array}\label{univ}
 \end{equation}
Here $u$ is a function on the 4-dimensional manifold $M$ with local coordinates $x, y, t, z$.
The characteristic variety of this system is the intersection of three quadrics,
$$
\begin{array}{c}
p_xp_t-p_y^2+u_{y}p_x^2-u_xp_xp_y=0,\\
p_xp_z-p_yp_t+u_{t}p_x^2-u_{x}p_xp_t=0,\\
 p_yp_z-p_t^2+u_{t}p_xp_y-u_{y}p_xp_t=0,
 \end{array}
$$
which specify a rational normal curve in  $\mathbb{P}T^*M$ parametrised as
$$p_x=1,\ p_y=\lambda-u_x,\ p_t=\lambda^2-u_x\lambda-u_y, \ p_z=\lambda^3-u_x\lambda^2-u_y\lambda-u_t,$$ so that
$$
\omega(\lambda)= dx+(\lambda-u_x) dy+(\lambda^2-u_x\lambda-u_y)dt+(\lambda^3-u_x\lambda^2-u_y\lambda-u_t)dz.
$$
Equations (\ref{univ})   are equivalent to the commutativity conditions of the following vector fields,
\begin{equation*}
\begin{array}{c}
\partial_y-(\lambda-u_x) \partial_x, \\
\partial_t-(\lambda^2 -u_x\lambda-u_y)\partial_x, \\
\partial_z-(\lambda^3-u_x\lambda^2-u_y\lambda-u_t)\partial_x,
\end{array}
\end{equation*}
which constitute a dispersionless Lax representation.
Note the absence of $\partial_{\lambda}$, which indicates a close similarity with the Veronese web
hierarchy. Integral manifolds of these vector fields provide a two-parameter family of $\alpha$-manifolds
of the corresponding $GL(2, \bbbr)$ structure.

\medskip

This has a straightforward higher-dimensional generalisation:
the equations are $${u_{i,j+1}-u_{i+1,j}+u_ju_{1,i}-u_iu_{1,j}=0, ~~~ 0<i<j<n};$$
the $GL(2, \bbbr)$ structure is given by $$\omega(\lambda)=\sum_{i=1}^n(\lambda^{i-1}-u_1\lambda^{i-2}-\dots-u_{i-1})\,dx^i;$$
the Lax representation is
$$X_i=\partial_{x^i}-(\lambda^{i-1}-u_1\lambda^{i-2}-\dots-u_{i-1})\partial_{x^1}, ~~~ 1<i\leq n.$$
Considered altogether, these equations form an integrable hierarchy.

\subsection{$GL(2, \bbbr)$ structures via Adler-Shabat triples}

Further examples of $GL(2, \bbbr)$ structures arise as characteristic varieties on solutions to triples of consistent 3D second-order PDEs  discussed by Adler and Shabat in \cite{Adler},
\begin{equation}
\begin{array}{c}
u_{23}=f(u_1, u_2, u_3, u_{12}, u_{13}),\\
u_{24}=g(u_1, u_2, u_4, u_{12}, u_{14}),\\
u_{34}=h(u_1, u_3, u_4, u_{13}, u_{14}),
\end{array}
\label{Adler}
\end{equation}
where $u$ is a function on the 4-dimensional manifold $M$ with local coordinates $x^1, \dots, x^4$. Note that system (\ref{Ver}) belongs to class (\ref{Adler}). As yet another example of this type let us consider the system
\begin{equation}
u_{23}=\frac{u_{12}-u_{13}}{u_2-u_3}, ~~~ u_{24}=\frac{u_{12}-u_{14}}{u_2-u_4}, ~~~ u_{34}=\frac{u_{13}-u_{14}}{u_3-u_4}.
\label{A}
\end{equation}
Its characteristic variety  is defined by a system of quadrics,
$$
p_2p_3=\frac{p_1p_2-p_1p_3}{u_2-u_3}, ~~~ p_2p_4=\frac{p_1p_2-p_1p_4}{u_2-u_4}, ~~~ p_3p_4=\frac{p_1p_3-p_1p_4}{u_3-u_4},
$$
which specify a rational normal curve in $\mathbb{P}T^*M$  parametrised as $p_1=1,\ p_i=\frac{1}{\lambda-u_i}$, so that
$$
\omega(\lambda)=dx^1+\frac{1}{\lambda-u_2}dx^2+\frac{1}{\lambda-u_3}dx^3+\frac{1}{\lambda-u_4}dx^4.
$$
System (\ref{A}) is equivalent to the conditions of commutativity of the following vector fields,
$$
\partial_{x^2}+\frac{1}{u_2-\lambda}\partial_{x^1}+\frac{u_{12}}{u_2-\lambda}\partial_{\lambda}, ~~~
\partial_{x^3}+\frac{1}{u_3-\lambda}\partial_{x^1}+\frac{u_{13}}{u_3-\lambda}\partial_{\lambda}, ~~~
\partial_{x^4}+\frac{1}{u_4-\lambda}\partial_{x^1}+\frac{u_{14}}{u_4-\lambda}\partial_{\lambda},
$$
note the explicit presence of $\partial_{\lambda}$. Projecting their integral manifolds from the extended
space $\hat M$ to $M$ we obtain a two-parameter family of $\alpha$-manifolds
of the corresponding $GL(2, \bbbr)$ structure.

\medskip

This has a straightforward higher-dimensional generalization:
the equations are $${(u_i-u_j)u_{ij}=u_{1i}-u_{1j}, ~~~ 1<i<j\leq n;}$$
the $GL(2, \bbbr)$ structure is given by $$\omega(\lambda)=dx^1+\sum_{i=2}^n\frac1{\lambda-u_i}\,dx^i;$$
the Lax representation is
$$X_i=\partial_{x^i}-\frac1{\lambda-u_i}\partial_{x^1}-\frac{u_{1i}}{\lambda-u_i}\partial_{\lambda}, ~~~ 1<i\leq n.$$
Considered altogether, these equations form an integrable hierarchy.

\section{General involutive $GL(2, \bbbr)$ structures}
\label{sec:gen}

In this section we demonstrate that general involutive $GL(2,\bbbr)$ structures are governed by a
dispersionless integrable hierarchy and derive the corresponding Lax system describing $\alpha$-manifolds.


\subsection{Parametrisation of involutive $GL(2, \bbbr)$ structures} \label{sec:alpha-par}

We begin by encoding all involutive structures in a simple ansatz.

 \begin{theorem}\label{Prop1}\po
Every involutive $GL(2, \bbbr)$ structure can be locally represented by formula (\ref{form}),
which upon clearing the denominators takes the form
 \begin{equation}\label{form-2}
\omega(\lambda)=\sum_{i=1}^n \Big[\prod_{j\ne i}\left(\lambda-\frac{u_j}{v_j}\right)\Big] u_i dx^i.
 \end{equation}
Here  $u$ and $v$ are functions of $(x^1, \dots, x^n)$ and subscripts denote partial derivatives: $u_i=u_{x^i}, \ v_i=v_{x^i}$. The functions $u$ and $v$  must satisfy a system of PDEs (\ref{un}),\,(\ref{vn}) coming from the integrability condition $d\omega(\lambda)\wedge \omega(\lambda)=0$.
 \end{theorem}

\centerline{\bf Proof:}

\medskip

Let (\ref{o}) be an involutive $GL(2, \bbbr)$ structure on $n$-dimensional manifold $M$.
It is easy to see that the space of $\alpha$-manifolds is at least 2-dimensional (in fact, it is parametrised by 1 arbitrary function of 1 variable, see Section \ref{sec:count}).
Choosing a 1-parameter family of $\alpha$-manifolds we obtain a (local) foliation  of $M$. This foliation consists of integral manifolds of an integrable distribution  $\omega(a)=0$ obtained by substituting $\lambda$  with some {\it function} $a$ on $M$.  We can thus set $\omega(a)= f d x$ for some functions $ f$ and $ x$. Let us now choose  $n$ different 1-parameter families of $\alpha$-manifolds that correspond to the choice of $n$ functions $a_i$ such that $\omega(a_i)= f_i d x^i$ (no summation). We will use $x^i$ as a local coordinate system on $M^n$. Note that although one can always set, say, $f_1=1$ by using conformal freedom in $\omega$, it is not always possible to eliminate all $f_i$ simultaneously. Taking into account that $\omega$ is polynomial (of degree $n-1$) in $\lambda$, the above conditions fix $\omega$ uniquely:
$$
\omega(\lambda)=\sum_{i=1}^n \Big[\prod_{j\ne i}\frac{\lambda-a_j}{a_i-a_j}\Big] f_i dx^i.
$$
Let us choose two extra 1-parameter families of $\alpha$-manifolds such   that $\omega(a_{n+1})= f_{n+1} du$ and $\omega(a_{n+2})= f_{n+2} dv$ (here $u, v$ are precisely the functions that will appear later in  formula (\ref{form})). Explicitly, this gives
 \begin{equation}
f_i \prod_{j\ne i}\frac{a_{n+1}-a_j}{a_i-a_j}=f_{n+1}u_i, ~~~
f_i \prod_{j\ne i}\frac{a_{n+2}-a_j}{a_i-a_j}=f_{n+2}v_i.
\label{ai}
 \end{equation}
The first of these relations allows one to rewrite $\omega$ as
 \begin{equation}
\omega(\lambda)=f_{n+1}\sum_{i=1}^n \Big[\prod_{j\ne i}\frac{\lambda-a_j}{a_{n+1}-a_j}\Big] u_i dx^i.
\label{omeg}
 \end{equation}
Taking the ratio of relations (\ref{ai}) we obtain
 $$
\prod_{j\ne i}\frac{a_{n+1}-a_j}{a_{n+2}-a_j}=\frac{f_{n+1}}{f_{n+2}}\frac{u_i}{v_i},
 $$
which is equivalent to
 $$
\frac{a_{n+2}-a_i}{a_{n+1}-a_i}=s \frac{u_i}{v_i}, ~~~ s=\frac{f_{n+1}}{f_{n+2}}\prod_{k=1}^n\frac{a_{n+2}-a_k}{a_{n+1}-a_k}.
 $$
Solving the last relation for $a_i$ and substituting the result into (\ref{omeg}) yields
 $$
\omega(\lambda)=f_{n+1}\sum_{i=1}^n \Big[\prod_{j\ne i}\frac{\lambda-a_{n+2}-s(\lambda-a_{n+1})\frac{u_j}{v_j}}{a_{n+1}-a_{n+2}}\Big] u_i dx^i.
 $$
Using the linear-fractional freedom in $\lambda$ (sending $a_{n+1}$ and $a_{n+2}$ to $\infty$ and $0$, respectively), as well as the conformal freedom in $\omega$, we can reduce the last expression to
form \eqref{form-2}.

Calculating the integrability condition $d\omega(\lambda)\wedge \omega(\lambda)=0$ (it is more convenient
to use \eqref{form} for this purpose) and collecting coefficients at $dx^i\wedge dx^j\wedge dx^k$ we obtain
\begin{equation}
\begin{array}{c}
\frac{\lambda-a_i}{u_i}\left(\frac{1}{\lambda-a_k}-\frac{1}{\lambda-a_j} \right)\lambda_i+
\frac{\lambda-a_j}{u_j}\left(\frac{1}{\lambda-a_i}-\frac{1}{\lambda-a_k} \right)\lambda_j+
\frac{\lambda-a_k}{u_k}\left(\frac{1}{\lambda-a_j}-\frac{1}{\lambda-a_i} \right)\lambda_k+S_{ijk}=0.
\end{array}
\label{lam}
\end{equation}
Here $\lambda_i=\lambda_{x^i}$ ($\lambda$ is viewed as a function of $x$), and
 $$
\begin{array}{c}
S_{ijk}=u_{ij}\frac{a_j-a_i}{u_iu_j}\left(\frac{\lambda}{\lambda-a_i}+\frac{\lambda}{\lambda-a_j} \right)+
u_{ik}\frac{a_i-a_k}{u_iu_k}\left(\frac{\lambda}{\lambda-a_i}+\frac{\lambda}{\lambda-a_k} \right)+
u_{jk}\frac{a_k-a_j}{u_ju_k}\left(\frac{\lambda}{\lambda-a_j}+\frac{\lambda}{\lambda-a_k} \right)\\
~~~~~~
-v_{ij}\frac{a_j-a_i}{u_iu_j}\left(\frac{\lambda a_i}{\lambda-a_i}+\frac{\lambda a_j}{\lambda-a_j} \right)
-v_{ik}\frac{a_i-a_k}{u_iu_k}\left(\frac{\lambda a_i}{\lambda-a_i}+\frac{\lambda a_k}{\lambda-a_k} \right)
-v_{jk}\frac{a_k-a_j}{u_ju_k}\left(\frac{\lambda a_j}{\lambda-a_j}+\frac{\lambda a_k}{\lambda-a_k} \right).
\end{array}
 $$
System  (\ref{un}),\,(\ref{vn}) governing general involutive $GL(2, \bbbr)$ structures results on  elimination of the derivatives of $\lambda$  from equations (\ref{lam}). This can be done as follows. Let us denote $T_{ijk}$ the left-hand side of (\ref{lam}). Taking 4 distinct indices $i\ne j\ne k\ne l$ one can verify that there are only two non-trivial linear combinations, namely
$$
T_{ikj}+T_{ijl}+T_{ilk}+T_{jkl}
$$
and
$$
\frac{1}{\lambda-a_l}T_{ikj}+\frac{1}{\lambda-a_k}T_{ijl}+\frac{1}{\lambda-a_j}T_{ilk}+\frac{1}{\lambda-a_i}T_{jkl},
$$
that do not contain derivatives of $\lambda$. The first linear combination is equal to zero identically, while the second combination vanishes (identically in $\lambda$) if and only if  relations (\ref{un}) and (\ref{vn}) are satisfied,
namely the following expression must vanish:
 $$
E_{ijkl}=\mathop{\mathfrak{S}}\limits_{(jkl)} (a_i-a_j)(a_k-a_l)\left( \frac{2u_{ij}-(a_i+a_j)v_{ij}}{u_iu_j}+ \frac{2u_{kl}-(a_k+a_l)v_{kl}}{u_ku_l} \right),
 $$
as well as  similar expressions obtained by interchanging $u$ and $v$,
 $$
F_{ijkl}=\mathop{\mathfrak{S}}\limits_{(jkl)}(b_i-b_j)(b_k-b_l)\left( \frac{2v_{ij}-(b_i+b_j)u_{ij}}{v_iv_j}+ \frac{2v_{kl}-(b_k+b_l)u_{kl}}{v_kv_l} \right),
 $$
recall that $a_i=\frac{u_i}{v_i}, \ b_i=\frac{v_i}{u_i}$. \hfill$\Box$

\medskip

Although  system (\ref{un}),\,(\ref{vn}) formally consists of $2\binom{n}{4}$ equations, only $2\binom{n-2}{2}$ of them are linearly independent. Indeed, we can restrict to equations $E_{12kl}=0$ and $F_{12kl}=0$  for $3\leq k<l\leq n$ since all other equations are their linear combinations: denoting $\a_{ij}=a_i-a_j$ we have
 \begin{equation}\label{EEEE}
\a_{12}E_{ijkl}=\a_{kl}E_{12ij}+\a_{jl}E_{12ki}+\a_{jk}E_{12il}+\a_{il}E_{12jk}+\a_{ik}E_{12lj}+\a_{ij}E_{12kl}
 \end{equation}
for all indices distinct (note that $\a_{ij}\neq0$ for $i\neq j$), and similarly for $F_{ijkl}$.

For $n=4$ system (\ref{un}),\,(\ref{vn}) is determined: it consists of 2 second-order PDEs for 2 functions $u$ and $v$
of 4 independent variables, so its general solution is parametrised by $4$ arbitrary functions of $3$ variables.
This gives an explicit confirmation of the result of \cite{Bryant} that modulo diffeomorphisms general involutive
$GL(2,\bbbr)$ structures in four dimensions  depend on $4$  functions of $3$ variables.
The case of general $n$ is more complicated because system (\ref{un}),\,(\ref{vn}) becomes overdetermined.

 \begin{theorem}\label{T2}\po
For every value of $n$,  the following holds:

\noindent {\bf (a)} The characteristic variety of system (\ref{un}),\,(\ref{vn}) is the tangential
variety of  rational normal curve (\ref{form}); it has degree $2n-4$. Rational normal curve (\ref{form}) can be recovered as the singular locus of the characteristic variety.

\noindent {\bf (b)} System (\ref{un}),\,(\ref{vn}) is in involution.

\noindent {\bf (c)} The general solution of system (\ref{un}),\,(\ref{vn}) depends on $2n-4$ functions of 3 variables (in the analytic or formal categories).
 \end{theorem}

\centerline{\bf Proof:}

\medskip

\noindent {\bf (a)} Let us parametrize rational normal curve (\ref{form}) as
 \begin{equation}\label{Cpl}
\lambda\mapsto [p_1:\dots:p_n]\in\mathbb{P}T^*M,\quad p_i=\frac{u_i}{\lambda-a_i}, \quad a_i=\frac{u_i}{v_i},
 \end{equation}
so that its  tangential variety is given by
 \begin{equation}\label{Cplm}
(\lambda,\mu)\mapsto [p_1:\dots:p_n]\in\mathbb{P}T^*M,\quad
 p_i=\frac{u_i}{\lambda-a_i}+\frac{u_i\mu}{(\lambda-a_i)^2}.
 \end{equation}
Let $E=E[u,v]$ and $F=F[u,v]$ be non-linear differential operators on the
left-hand sides of (\ref{un}) and (\ref{vn}). The symbol of the system $\mathcal{E}=\{E=0,\ F=0\}$ is given by the matrix
 \begin{equation}\label{ME}
\ell_\mathcal{E}(p)=\begin{bmatrix}\ell_E^u(p) & \ell_E^v(p)\\ \ell_F^u(p) & \ell_F^v(p)\end{bmatrix},
 \end{equation}
where $\ell_E^u(p)=\sum\limits_{a\le b}\frac{\partial E}{\partial u_{ab}}p_ap_b$ is the
symbol of $u$-linearization of $E$, etc.
As  noted after Theorem \ref{Prop1}, $E=(E_{ijkl})$ has $\binom{n-2}{2}$ independent components, and similarly for
$F=(F_{ijkl})$, so that the matrix $\ell_\mathcal{E}$ is of  the size $2\binom{n-2}{2}\times 2$.
The characteristic variety is defined by the formula
 $$
\op{Char}(\mathcal{E})=\{[p]\in\mathbb{P} T^*M:\op{rank}\bigl(\ell_\mathcal{E}(p)\bigr)<2\}.
 $$
From (\ref{un}) we have
 $$
\ell_{E_{ijkl}}^u(p)=2\mathop{\mathfrak{S}}\limits_{(jkl)}(a_i-a_j)(a_k-a_l)\Bigl(\frac{p_ip_j}{u_iu_j}
+\frac{p_kp_l}{u_ku_l}\Bigr).
 $$
This expression vanishes if we substitute $p$ from \eqref{Cpl}. Similarly, all  other components
$\ell_{E_{ijkl}}^v(p)$, $\ell_{F_{ijkl}}^u(p)$, $\ell_{F_{ijkl}}^v(p)$ of the
symbolic matrix vanish, and we conclude that $\ell_\mathcal{E}(p)=0$ modulo \eqref{Cpl}.

For the tangential variety \eqref{Cplm},  the entries of $\ell_\mathcal{E}(p)$ do not vanish identically,
however, a straightforward computation shows that independently of $(ijkl)$ we get
 $$
\lambda\,\ell_{E_{ijkl}}^u(p)+\ell_{E_{ijkl}}^v(p)=0 \text{~~ and ~~}
\lambda\,\ell_{F_{ijkl}}^u(p)+\ell_{F_{ijkl}}^v(p)=0,
 $$
and these identities characterise (\ref{Cplm}). Thus, all columns of $\ell_\mathcal{E}(p)$ are proportional
whenever $p$ satisfies \eqref{Cplm}, and
$\op{rank}\bigl(\ell_\mathcal{E}(p)\bigr)=1$ unless $p$ belongs to the rational normal curve (in which case
we have $\op{rank}\bigl(\ell_\mathcal{E}(p)\bigr)=0$). Finally, for a rational normal curve of degree $n-1$, the degree of its tangential variety equals $2n-4$. This  variety  is known to be   generated by quartics \cite{Harris}.

\medskip

\noindent {\bf (b)} System $\mathcal{E}=\{E_{12kl}=0,F_{12kl}=0\}$ given by (\ref{un}), (\ref{vn}) is involutive iff
its compatibility conditions are identically satisfied modulo $\E$. A long computation, which we present in
Appendix \ref{App.A}, shows that these conditions are numerated by 5-tuples of distinct indices $(12ijk)$
where $2<i<j<k\leq n$. More precisely, the compatibility conditions corresponding to any such 5-tuple are
first-order differential operators applied to $E_{12ij},E_{12jk},E_{12ki}$ and $F_{12ij},F_{12jk},F_{12ki}$,
and involving only differentiations by variables $x^1,x^2,x^i,x^j,x^k$.
There are four compatibility conditions for each 5-tuple $(12ijk)$.

Thus it suffices to check compatibility for $n=5$ to conclude it for  general $n$.
For $n=5$ the resolution from Appendix \ref{App.A} becomes a short exact sequence
$\mathcal{R}^{2} \stackrel{\ell_\mathcal{E}}\longrightarrow\mathcal{R}^{6}
\stackrel{\mathcal{C}_\E}\longrightarrow\mathcal{R}^{4}$,
where $\mathcal{R}=\R[p_1,\dots,p_n]$ is the algebra of homogeneous polynomials on $T^*M$
and $\mathcal{C}_\E$ is the compatibility operator.
From this we read off the 4 compatibility conditions.  A direct verification
(using symbolic computations in Maple) shows that they are satisfied. This implies the involutivity.

\medskip

\noindent {\bf (c)} By a classical result going back to Cartan the general local solution of an involutive PDE system
$\E$ depends on $d$ arbitrary functions of $m$ variables where the numbers $d$ (formal rank) and $m$ (formal dimension)
can be read off the Cartan characters characterising involutivity. The result is formal, but it also holds in the
analytic category due to the Cartan-K\"ahler theorem. Serre reformulated this criterion in homological terms,
relating the numbers $d,m$ to the Hilbert function of the symbolic module. Since the characteristic variety
is the support of this module, these numbers can be read off the geometry of this variety and the sheaf
$\op{ker}(\ell_\E)$ over $\op{Char}(\E)$.

We refer to \cite{BCG, KL} for a modern exposition of these results. In the case when $\op{Char}(\E)$ is irreducible
the number $m$ is the affine dimension of this variety, while $d$ is its degree multiplied by the rank of the
sheaf $\op{ker}\bigl(\ell_\E(p)\bigr)$ at generic point $p\in\op{Char}(\E)$.
Since system (\ref{un}),\,(\ref{vn}) is in involution and its characteristic variety has
affine dimension $m=3$, degree $d=2n-4$ and the kernel sheaf of dimension $2-\op{rank}\bigl(\ell_\E(p)\bigr)=1$ at
any point $p\in\op{Char}(\E)$ that belongs to \eqref{Cplm} with $\mu\neq0$,
the general solution  depends on $2n-4$ arbitrary functions of $3$ variables.
\hfill$\Box$

 \medskip

 \noindent{\bf Remark 1.}
The system $\E$ can be represented in a simple parametric form ($1\leq i< j\leq n$)
 $$
\begin{array}{c}
\frac{2u_{ij}-(a_i+a_j)v_{ij}}{u_iu_j}=r_i+r_j+\sum_{k=3}^{n-1} l_k\frac{a_i^k-a_j^k}{a_i-a_j}, ~~~
\frac{2v_{ij}-(b_i+b_j)u_{ij}}{v_iv_j}=s_i+s_j+\sum_{k=3}^{n-1} m_k\frac{b_i^k-b_j^k}{b_i-b_j},
\end{array}
 $$
This system has $n(n-1)$ equations
and \mbox{$4n-6$} parameters $r_1,\dots,r_n$, $s_1,\dots,s_n$, $l_3,\dots l_{n-1}$, $m_3,\dots, m_{n-1}$.
Elimination of these parameters yields $(n-2)(n-3)$ equations (\ref{un}),\,(\ref{vn}).


\subsection{Integrability of involutive $GL(2, \bbbr)$ structures} \label{sec:alpha-int}

 \begin{theorem}\label{T1}\po
For every $n$, system  (\ref{un}),\,(\ref{vn}) is integrable via
a dispersionless Lax representation in parameter-dependent vector fields.
Letting $n\to \infty$ we obtain the corresponding dispersionless integrable hierarchy.
 \end{theorem}


\centerline{\bf Proof:}

\medskip

Let us associate with equations (\ref{lam}) the following family of $\lambda$-dependent vector fields,
 $$
\begin{array}{c}
V_{ijk}=\frac{\lambda-a_i}{u_i}\left(\frac{1}{\lambda-a_k}-\frac{1}{\lambda-a_j} \right)\partial_{x^i}+
\frac{\lambda-a_j}{u_j}\left(\frac{1}{\lambda-a_i}-\frac{1}{\lambda-a_k} \right)\partial_{x^j}+
\frac{\lambda-a_k}{u_k}\left(\frac{1}{\lambda-a_j}-\frac{1}{\lambda-a_i} \right)\partial_{x^k}-S_{ijk}\partial_{\lambda},
\end{array}
 $$
which live in the extended space $\hat M$ with coordinates $x^1, \dots, x^n, \lambda$. These vector fields generate a distribution $V=span \langle V_{ijk}\rangle$ in $T\hat M$ of dimension $n-2$.
Indeed, the identities noted in the proof of Theorem \ref{Prop1} for $T_{ijk}$ hold for $V_{ijk}$,
so these latter vector fields are expressed as linear combinations of $V_{12l}$ for $3\leq l\leq n$. This, in particular, implies that modulo (\ref{un}),\,(\ref{vn}) there are only $n-2$ linearly independent relations (\ref{lam}).

The geometry behind system  (\ref{lam}) and the distribution $V$ is as follows. Consider a hypersurface $H$ in $\hat M$
defined explicitly as  $\lambda=\lambda(x^1, \dots, x^n)$. Then the distribution $V$ is tangential to  $H$ if and only if
the function $\lambda(x^1, \dots, x^n)$ solves system (\ref{lam}). Thus system (\ref{lam}) is compatible if and only if
the associated distribution $V$ is involutive. In this case the general solution of system (\ref{lam}) depends on
1 arbitrary function of 2 variables: there exists  a 3-parametric family of integral manifolds of $V$, and
a generic hypersurface $H\subset \hat M$ with $V|_H\subset TH$ is formed by a 2-parametric subfamily of integral
manifolds of $V$, whence the functional freedom.

Direct calculation based on the Frobenius theorem shows that by virtue of equations (\ref{un}),\,(\ref{vn})
the distribution $V$ is involutive.
Thus, $\lambda$-dependent vector fields $V_{ijk}$ constitute a dispersionless Lax representation for  system (\ref{un}),\,(\ref{vn}). Projecting integral manifolds of $V$ from $\hat M$ to $M$ we obtain a $3$-parameter family of codimension 2 submanifolds of $M$. Tangent spaces to these submanifolds are $(n-2)$-dimensional osculating spaces of the dual curve $\tilde \omega (\lambda)$. Indeed, the distribution $V$ is annihilated by the (pulled-back) 1-forms $\omega(\lambda)$ and $\omega'(\lambda)$.

Equations (\ref{un}),\,(\ref{vn}) for $u$ and $v$ are organised in pairs, each pair involving 4 independent variables
indexed from 1 to $n$. As $n$ grows, the collection of PDEs is nested and compatible.
Ultimately when $n\to \infty$  we obtain the corresponding dispersionless hierarchy.
\hfill$\Box$

\medskip

In the context of the general heavenly hierarchy, similar Lax equations appeared recently in \cite{Bogdanov}.  A  modification of the inverse scattering transform for Lax equations in parameter-dependent vector fields was developed in \cite{Man-San}.

\medskip

\noindent{\bf Remark 2.}
System (\ref{un}),\,(\ref{vn}) governing general involutive  $GL(2, \bbbr)$ structures can be viewed as a generalisation of the Veronese web hierarchy. Indeed, the Veronese web hierarchy results upon setting $v_i=\frac{1}{c_i}qu_i$, where $c_i$ are constants and $q$ is some function. Then the reparametrisation $\lambda \to
\lambda/q$ identifies $GL(2, \bbbr)$ structure (\ref{form}) with (\ref{Vero}) (up to unessential conformal factor $q$), so that system (\ref{un}),\,(\ref{vn}) reduces to equations (\ref{Ver}) of the Veronese web hierarchy.
Note that reductions of the general system (\ref{un}),\,(\ref{vn}) to other examples of Sect. \ref{sec:ex} (say, the dKP hierarchy) are far more complicated,  requiring highly transcendental nonlocal changes of the independent variables $x^i$ and the dependent variables $u, v$.
Indeed, although  the  coordinate planes $x^i=const$ constitute $\alpha$-manifolds for  $GL(2, \bbbr)$ structure (\ref{form}), this is not the case for the dKP hierarchy.

Another class of (translationally non-invariant) integrable deformations of the Veronese web hierarchy was considered recently in \cite{KP}: the corresponding Lax equations  do not however contain $\partial_\lambda$, and are specifically 3-dimensional.

\medskip

\noindent{\bf Remark 3.} For $n=4$ there exists a unique torsion-free $GL(2, \bbbr)$ connection associated with $GL(2, \bbbr)$ structure (\ref{form}). It can be parametrised as
$$
\begin{array}{c}
\Gamma^i_{jk}=\frac{u_i^2u_ju_k}{(a_i-a_j)(a_i-a_k)}\psi_i, ~~~
\Gamma^i_{jj}=\frac{u_i^2u_j^2}{(a_i-a_j)^2}\psi_i, ~~~ \Gamma^i_{ij}=\Gamma^i_{ji}=\frac{u_iu_j}{a_i-a_j}\phi_i, ~~~ \Gamma^i_{ii}=\rho_i,\\
\end{array}
$$
where   $i, j, k\in \{1, \dots, 4\}$ are pairwise distinct indices,  and the quantities $\psi_i, \phi_i, \rho_i$ are yet to be determined from the following linear system with extra parameters $s_j, \tilde s_j$ to be eliminated.
 $$
\frac{u_{ij}}{u_iu_j}-\sum_k \Gamma^k_{ij}\frac{u_k}{u_iu_j}=s_2a_ia_j+s_1(a_i+a_j)+s_0,\quad
\frac{v_{ij}}{v_iv_j}-\sum_k \Gamma^k_{ij}\frac{v_k}{v_iv_j}=\tilde s_2b_ib_j+\tilde s_1(b_i+b_j)+\tilde s_0.
 $$
\comm{
\begin{equation}
\begin{array}{c}
\frac{u_{ij}}{u_iu_j}-\sum_k \Gamma^k_{ij}\frac{u_k}{u_iu_j}=s_2a_ia_j+s_1(a_i+a_j)+s_0,\\
\ \\
\frac{v_{ij}}{v_iv_j}-\sum_k \Gamma^k_{ij}\frac{v_k}{v_iv_j}=\tilde s_2b_ib_j+\tilde s_1(b_i+b_j)+\tilde s_0,
\end{array}
\label{symcon}
\end{equation}
 }%
This system contains 20 linear equations for the 18 unknowns $\psi_i, \phi_i, \rho_i, s_j, \tilde s_j$.
These equations are consistent modulo (\ref{un}),\,(\ref{vn}), and lead to a unique torsion-free $GL(2, \bbbr)$ connection.

\subsection{Counting $\alpha$-manifolds} \label{sec:alpha-num}
\label{sec:count}

The disperionless Lax representation provides a two-parametric family of $\alpha$-manifolds.
The totality of all $\alpha$-manifolds is bigger.

 \begin{prop}\label{Prop3}\po
For an involutive $GL(2, \bbbr)$ structure, its local $\alpha$-manifolds
are parametrised by 1 function of 1 variable.
 \end{prop}

\centerline{\bf Proof:}

\medskip

Let us invoke a relation with ordinary differential equations having all W\"unschmann invariants zero,
see \cite{Krynski} for details (recall that all involutive structures arise on solution spaces of such ODEs).
An ODE $\mathcal{E}$ of order $n$ is given
by a submanifold $x_n=F(t,x_0,x_1,\dots,x_{n-1})$ in the jet-space
$J^n=\R^{n+2}(t,x_0,\dots,x_n)$, and $\mathcal{E}$ is diffeomorphic (via the jet-projection)
to the jet-space $J^{n-1}$.
The solution space $M^n$ is identified with the space of  integral curves of the field
$X_F=\p_t+x_1\p_0+\dots+x_{n-1}\p_{n-2}+F\p_{n-1}$, where  $\p_i=\p_{x_i}$ and $F=F(t,x_0,x_1,\dots,x_{n-1})$.

Denote by $\pi:J^{n-1}\to M=J^{n-1}/X_F$ the projection
(since the construction is local, this quotient exists, and is non-singular),
and let $\mathcal{D}_{n-1}=\langle\p_1,\dots,\p_{n-1}\rangle$ be the vertical distribution in $J^{n-1}$ with respect to the projection of $J^{n-1}$ to $J^0=\R^2(t,x_0)$.
The family of hyperplanes $\pi_*\mathcal{D}_{n-1}\subset TM$ parametrised by the coordinate
$\lambda=t$ along integral curves of $X_F$ coincides with $\alpha$-hyperplanes of a $GL(2, \bbbr)$ structure on $M$ provided the W\"unschmann invariants vanish.

Thus $\alpha$-manifolds are projections of integral manifolds of  (maximal possible) dimension $n-1$
for the (non-holonomic) distribution
 $$
\mathcal{D}_{n}=\pi_*^{-1}\pi_*(\mathcal{D}_{n-1})=\langle X_F,\p_1,\dots,\p_{n-1}\rangle=
\langle\p_t+x_1\p_0,\p_1,\dots,\p_{n-1}\rangle.
 $$
This distribution has rank $n$ and possesses a sub-distribution of Cauchy characteristics of rank $n-2$ given by $Ch(\mathcal{D}_{n})=\langle \p_2,\dots,\p_{n-1}\rangle$.
Consequently, integral manifolds of $\mathcal{D}_{n}$  are foliated by the Cauchy characteristics, and therefore coincide with vertical lifts of  Legendrian curves of the standard contact structure on the quotient $J^1=J^{n-1}/Ch(\mathcal{D}_{n})$.

Note that generic Legendrian curves in $J^1=\R^3(t,x_0,x_1)$
are uniquely determined by their projection to the plane $J^0=\R^2(t,x_0)$;
the curves whose projections degenerate to a point correspond to the standard two-parameter family of $\alpha$-manifolds.
Since curves in the plane are parametrised by 1 function of 1 variable, the claim follows.
\hfill$\Box$

\medskip



\noindent{\bf Remark 4.} By a theorem of Sophus Lie a system of PDEs with the general solution depending on
1 function of 1 variable is solvable via ODEs \cite{SL,BK1L}.
Thus $\alpha$-manifolds of any involutive $GL(2, \bbbr)$ structure can be found as solutions to a system of ODEs.

\subsection{Equivalent definitions of involutivity}
\label{sec:ext}



 \begin{prop}\label{Prop+}\po
For a $GL(2,\bbbr)$ structure, the definitions of involutivity in the sense of Bryant \cite{Bryant2}
and $\alpha$-integrability in the sense of Krynski \cite{Krynski} are equivalent.
 \end{prop}

\centerline{\bf Proof:}

\medskip

Consider a manifold $M^n$, the associated  contact manifold $\mathbb{P}T^*M$ of
dimension $2n-1$ with the contact distribution $\mathcal{C}_M$,
and a submanifold $\mathcal{Z}\subset \mathbb{P}T^*M$ of dimension $n+1$ that corresponds to a $GL(2, \bbbr)$ structure on $M$. We have $\op{dim}(T\mathcal{Z}\cap \mathcal{C}_M)=n$.
Since the projection $\pi:\mathbb{P}T^*M\to M$ is surjective on $\mathcal{Z}$,
the intersection $\mathcal{Z}_x=\mathcal{Z}\cap\pi^{-1}(x)\subset \mathbb{P}T_x^*M$
is a curve (rational normal curve) for each $x\in M$. For $p\in\mathcal{Z}_x$  we have
$d_p\pi(T\mathcal{Z}\cap \mathcal{C}_M)=p^\perp\subset T_xM$,
$p^\perp\simeq T\mathcal{Z}\cap \mathcal{C}_M/T_p\mathcal{Z}_x$.

Denote the contact form by $\omega$. Then $\mathcal{Z}$ is {\em involutive} in the sense of \cite{Bryant2}
if $\omega\wedge(d\omega)^2|_{\mathcal{Z}}=0$ $\Leftrightarrow$
$(d\omega|_{T\mathcal{Z}\cap \mathcal{C}_M})^2=0$
(and $d\omega|_{T\mathcal{Z}\cap \mathcal{C}_M}\neq0$ for dimensional reasons),
whence for the subbundle
$\Pi_M=\op{Ker}(d\omega|_{T\mathcal{Z}\cap \mathcal{C}_M})$ we have $\op{rank}\Pi_M=n-2$.
The local quotient $(S_M,D_M)=(\mathcal{Z},T\mathcal{Z}\cap \mathcal{C}_M)/\Pi_M$ is a
3-dimensional contact manifold. Denoting the projection by $\rho:\mathcal{Z}\to S_M$,
the corresponding  $\alpha$-manifolds can be represented in the form $\rho^{-1}(L)$
where $L\subset S_M$ is a Legendrian curve with respect to $D_M$ (compare with the proof of
Proposition \ref{Prop3} from Section \ref{sec:count}).

Conversely, if for every $x\in M$, $p\in\mathcal{Z}_x$ there exists an $\alpha$-manifold tangent to
$p^\perp\subset T_xM$, then the restriction of the canonical conformally symplectic form $[d\omega]$
to $T\mathcal{Z}\cap \mathcal{C}_M$ has rank 2, so that $\alpha$-integrability
implies involutivity in the sense of \cite{Bryant2}.


\section{Concluding remarks}

We  conclude with two general comments.

\begin{itemize}

\item It was demonstrated that  involutive $GL(2, \bbbr)$ structures in 4D or, equivalently, torsion-free affine connections with the irreducible $GL(2, \bbbr)$ holonomy, are governed by a dispersionless integrable system. It would be interesting to understand which special holonomies lead to nonlinear PDEs that are either explicitly solvable/linearisable, or belong to the class of  integrable systems.

\item Interesting generalisations of involutive $GL(2, \bbbr)$ structures arise in the context of integrable hierarchies whose characteristic varieties are {\it elliptic curves}. For instance, the first two  equations of the dispersionless Pfaff-Toda hierarchy \cite{Takasaki}  are of the form (see \cite{Z2})
 \begin{gather*}
e^{F_{xx}}F_{xt}=e^{F_{yy}}F_{yz},\\
F_{zt}=2e^{F_{xx}+F_{yy}}\sinh (2F_{xy}).
 \end{gather*}
Here $F$ is a function on the 4-dimensional manifold $M$ with coordinates $x, y, t, z$. The
characteristic variety of this system is a complete intersection of two quadrics in $\mathbb{P}^3$:
 \begin{gather*}
e^{F_{xx}}p_xp_t+e^{F_{xx}}F_{xt}p_x^2=e^{F_{yy}}p_yp_z+e^{F_{yy}}F_{yz}p_y^2, \\
p_zp_t=e^{F_{xx}+F_{yy}}(e^{2F_{xy}}(p_x+p_y)^2-e^{-2F_{xy}}(p_x-p_y)^2).
 \end{gather*}
This specifies a field of elliptic curves in the projectivised cotangent bundle $\mathbb{P} T^*M$, recall that the genus $g$ of a nonsingular complete intersection of two nonsingular surfaces of degrees $d, e$ in $\mathbb{P}^3$ equals $g=\frac{1}{2}de(d+e-4)+1$, see e.g. \cite{Hartshorne}, Chapter 2, exercise 8.4 (g). For $d=e=2$ this gives $g=1$.
The geometry of such structures is yet unclear, primarily due to the lack of a  naturally adapted connection (analogous to $GL(2, \bbbr)$ connection) compatible with the above family of elliptic curves in the spirit of (\ref{K}) (indeed, any such connection would automatically preserve all scalar differential invariants of the curves, however, their  $j$-invariants are non-constant).
\end{itemize}

\appendix

\section{Compatibility conditions via free resolutions}\label{App.A}

In this section we explain how techniques from commutative algebra can be used to effectively
compute compatibility conditions of overdetermined systems of PDEs. Then we apply this to our
overdetermined system \eqref{un}-\eqref{vn} encoding involutive $GL(2,\R)$ structures.

We refer to \cite{KL,KL2} for details on this approach to involutivity of overdetermined systems of PDEs,
and for a background on jet machinery in the formal theory of differential equations.

\subsection{Projective resolutions and linear differential operators}

Let us first consider the case of a linear system $\E$ of PDEs given by a $k$-th order differential operator
$\Delta:\Gamma(\pi)\to\Gamma(\nu)$ on sections of vector bundles $\pi,\nu$ over $M$. Such an operator corresponds to
a morphism of vector bundles $\psi_k^\Delta:J^k\pi\to\nu$ with jet-prolongations $\psi_{k+i}^\Delta:J^{k+i}\pi\to J^i\nu$.
Then $\E_{k+i}=\op{Ker}(\psi_{k+i}^\Delta)$ for $i\ge0$ and $\E_l=J^l\pi$ for $0\leq l<k$. The bundle
$\E_\infty$ is the projective limit of $\E_k$ with respect to projections $\pi_{i+1,i}:\E_{i+1}\to\E_i$.

The dual bundle $\E^*=\{\E_k^*\}$ allows to characterise involutivity as follows: the system $\E$ is compatible
(involutive) iff $\Gamma(\E_k^*)$ are projective $C^\infty(M)$-modules and $\pi_{i+1,i}^*$ are injective.
Compatibility complex is related to projective resolution of the module $\Gamma(\E_\infty^*)$, but it
is more convenient to construct this at the symbolic level.

At a point $x\in M$ the symbol sequence of $\E$ is $g_k=\op{Ker}(d\pi_{k,k-1}:T_x\E_k\to T_x\E_{k-1})\subset
S^kT^*_xM\ot\pi$.
The dual (over $\R$) sequence determines the module $\mathcal{M}_\E=\oplus g_k^*$ over the algebra
$\mathcal{R}=ST=\oplus_{i=0}^\infty S^iT_xM$ of homogeneous polynomials on $T^*_xM$, called the symbolic module of $\E$.

Since localisation of a projective module is free, we can construct a minimal free resolution of this module,
where $\sigma_\Delta$ is the symbol of $\Delta$, the dual of which defines relations among the generators of
$\mathcal{M}_\E$, and $\psi^*$ is the first syzygy (we use $*$ for further convenience):
 \begin{equation}\label{Rcomplex}
\dots\to \mathcal{R}\otimes\varpi^*\stackrel{\psi^*}\longrightarrow
\mathcal{R}\otimes\nu^*\stackrel{\sigma_\Delta^*}\longrightarrow
\mathcal{R}\otimes\pi^*\longrightarrow \mathcal{M}_\E\to 0.
 \end{equation}
Applying to this the functor $*=\op{Hom}_\R(\cdot,\R)$ we get the following exact sequence
 $$
0\to g\hookrightarrow ST^*\otimes\pi\stackrel{\sigma_\Delta}\longrightarrow
ST^*\otimes\nu\stackrel{\psi}\longrightarrow ST^*\otimes\varpi\to\dots
 $$
from which we obtain the compatibility condition for $\E=\{\Delta=0\}$ as follows.
Let $\Psi\in\op{Diff}(\nu,\varpi)$ be a differential operator with the symbol $\psi$ at $x$.
Then the compatibility is $\Psi\circ\Delta|_\E=0$.

More specifically, if the operator $\Delta$ has order $k$ and $\Psi$ has order $m$ (we consider the simplest case
when we have only one order), then $\Psi\circ\Delta$ has order $\leq k+m-1$ and it should be in the differential
ideal of $\E$, so that $\Psi\circ\Delta=\Xi\circ\Delta$ for a differential operator $\Xi$ of order $<m$.
Modification $\Psi\mapsto\Psi'=\Psi-\Xi$ does not change the symbol and we get what is called the differential syzygy:
 \begin{equation}\label{DSyz}
\Psi'\circ\Delta=0.
 \end{equation}
This is how algebraic syzygy determines compatibility conditions in the linear case.

For nonlinear equations, apply the linearisation operator on a solution instead of $\Delta$. Its symbol again
leads to a syzygy, from which we deduce compatibility operators; in this case however $\Psi$ is an operator
in total derivatives. Differential syzygy \eqref{DSyz}, considered as a differential corollary of $\E$,
yields the complete compatibility condition for this system.

\subsection{Application to involutive $GL(2,\R)$ structures}

Let us indicate how to construct the syzygy $\psi$ corresponding to system \eqref{un}-\eqref{vn}.
In this case the order of the system is $k=2$ and the order of the syzygy will be $m=1$.

Recall that at fixed point $x\in M$ we denote $\mathcal{R}=\R[p_1,\dots,p_n]=ST_xM=\oplus_{k=0}^\infty S^kT_xM$
the algebra of homogeneous polynomials on $T_x^*M$. Denote also $\mathcal{R}^q=\mathcal{R}\otimes_\R\R^q$.

The symbol $\ell_\E$ of the nonlinear vector-operator defining $\E$ is given by matrix \eqref{ME}, and
in new coordinates $\xi_i=\frac{p_i}{u_i}$ on $T^*_xM$ it has components
 \begin{alignat*}{1}
\ell^u_{E_{ijkl}}(\xi) &=
2\mathop{\mathfrak{S}}\limits_{(jkl)}(a_i-a_j)(a_k-a_l)\bigl(\xi_i\xi_j+\xi_k\xi_l\bigr),
\\
\ell^v_{E_{ijkl}}(\xi) &=
-\mathop{\mathfrak{S}}\limits_{(jkl)}(a_i-a_j)(a_k-a_l)\bigl((a_i+a_j)\xi_i\xi_j+(a_k+a_l)\xi_k\xi_l\bigr),
\\
\ell^u_{F_{ijkl}}(\xi) &=
-\mathop{\mathfrak{S}}\limits_{(jkl)}\frac{(a_i-a_j)(a_k-a_l)}{a_ia_ja_ka_l}\bigl((a_i+a_j)\xi_i\xi_j+(a_k+a_l)\xi_k\xi_l\bigr),
\\
\ell^v_{F_{ijkl}}(\xi) &=
2\mathop{\mathfrak{S}}\limits_{(jkl)}(a_i-a_j)(a_k-a_l)\Bigl(\frac{\xi_i\xi_j}{a_ka_l}+\frac{\xi_k\xi_l}{a_ia_j}\Bigr),
 \end{alignat*}
in the basis $e_u,e_v$ of $\mathcal{R}^2$ and basis $e_{E_{ijkl}}$, $e_{F_{ijkl}}$ of $\mathcal{R}^{2\binom{n-2}2}$,
where due to relations \eqref{EEEE} we restrict to indices $i=1,j=2$, $2<k<l\leq n$.
This means that the homomorphism $\ell_\E$ maps $f(\xi)e_u$ to
$f(\xi)\sum_{k<l}(\ell^u_{E_{12kl}}(\xi)e_{E_{12kl}}+\ell^u_{F_{12kl}}(\xi)e_{F_{12kl}})$
and similarly for $h(\xi)e_v$.

Now we resolve $\ell_\E$ by a homomorphism $\mathcal{C}=\mathcal{C}_\E$.
For $w=\sum_{i<j}(w_{E_{12ij}}e_{E_{12ij}}+w_{F_{12ij}}e_{F_{12ij}})$ the image
$\mathcal{C}(\xi)(w)$ has the following components ($2<i<j<k\leq n$):
 \begin{alignat*}{1}
\mathcal{C}^I_{ijk}=&
\mathop{\mathfrak{S}}\limits_{(ijk)}\bigl((a_2-a_k)\xi_1+(a_k-a_1)\xi_2+(a_1-a_2)\xi_k\bigr)w_{E_{12ij}}, \\
\mathcal{C}^{II}_{ijk}=&
\mathop{\mathfrak{S}}\limits_{(ijk)}\Bigl[
\bigl((a_1-a_2)(a_2-a_k)a_1\xi_1+(a_2-a_1)(a_1-a_k)a_2\xi_2 \\
&+((a_2-a_k)^2a_1+(a_1-a_k)^2a_2)\xi_k\bigr)w_{E_{12ij}}+2a_1a_2a_ia_j(a_1-a_k)(a_2-a_k)\xi_kw_{F_{12ij}}\Bigr], \\
\mathcal{C}^{III}_{ijk}=&
\mathop{\mathfrak{S}}\limits_{(ijk)}\Bigl[2(a_1-a_k)(a_2-a_k)\xi_kw_{E_{12ij}}
+\bigl((a_1-a_2)(a_2-a_k)a_1\xi_1 \\
&+(a_2-a_1)(a_1-a_k)a_2\xi_2+((a_2-a_k)^2a_1+(a_1-a_k)^2a_2)\xi_k\bigr)a_ia_jw_{F_{12ij}}\Bigr], \\
\mathcal{C}^{IV}_{ijk}=&
\mathop{\mathfrak{S}}\limits_{(ijk)}
\bigl((a_2-a_k)a_1^2\xi_1+(a_k-a_1)a_2^2\xi_2+(a_1-a_2)a_k^2\xi_k\bigr)a_ia_jw_{F_{12ij}}.
 \end{alignat*}

One verifies that with these homomorphisms the following sequence is exact:
 \begin{equation}\label{264n}
\mathcal{R}^2\stackrel{\ell_\E}\longrightarrow\mathcal{R}^{2\binom{n-2}2}
\stackrel{\mathcal{C}_\E}\longrightarrow\mathcal{R}^{4\binom{n-2}3}.
 \end{equation}
In other words, $\mathcal{C}_\E$ is the first syzygy for the module
$\mathcal{M}_\E^\star=\op{Ker}(\ell_\E)=\op{Hom}_\mathcal{R}(\mathcal{M}_\E,\mathcal{R})$.
Therefore, the differential syzygies \eqref{DSyz} for $\E$ given by system \eqref{un}-\eqref{vn}
are enumerated by 5 different indices $(12ijk)$, $2<i<j<k\leq n$.
Consequently to verify compatibility conditions for each of these 5-tuples
one can work in the corresponding 5-dimensional space,
and this justifies the key argument used in the proof of part (b) of Theorem \ref{T2}.

\subsection{Constructing the minimal free resolution}

The higher syzygies resolve the dual $\mathcal{M}_\E^\star$ of the symbolic module as follows.
Let us recall a construction from the commutative algebra adapted to our situation. For a homomorphism
$\varphi:\mathcal{R}^{n-2}\to\mathcal{R}^2$ the following sequence is known as the Eagon-Northcott complex
 \cite[Appendix A2]{E} (all tensor products are over $\mathcal{R}$, and $\star$ is the dualisation over $\mathcal{R}$)
 $$
\dots\to S^3\mathcal{R}^{\star2}\otimes\Lambda^5\mathcal{R}^{n-2}\stackrel{\partial}\longrightarrow
S^2\mathcal{R}^{\star2}\otimes\Lambda^4\mathcal{R}^{n-2}\stackrel{\partial}\longrightarrow
\mathcal{R}^{\star2}\otimes\Lambda^3\mathcal{R}^{n-2}\stackrel{\partial}\longrightarrow
\Lambda^2\mathcal{R}^{n-2}\stackrel{\epsilon}\longrightarrow\mathcal{R}.
 $$
Here $\epsilon=\La^2\varphi$ and the differential $\partial$ is the following composition,
in which $\delta$ is the Spencer differential and $\divideontimes$ is the Hodge dual via a volume form:
 \begin{multline*}
S^{d+1}\mathcal{R}^{\star2}\otimes\Lambda^{d+3}\mathcal{R}^{n-2}
\stackrel{\delta\otimes\divideontimes}\longrightarrow
S^d\mathcal{R}^{\star2}\otimes\mathcal{R}^{\star2}\otimes\Lambda^{n-d-5}\mathcal{R}^{\star(n-2)}
\stackrel{1\otimes\varphi^\star\otimes1}\longrightarrow\\
S^d\mathcal{R}^{\star2}\otimes\mathcal{R}^{\star(n-2)}\otimes\Lambda^{n-d-5}\mathcal{R}^{\star(n-2)}
\stackrel{1\otimes\wedge}\longrightarrow
S^d\mathcal{R}^{\star2}\otimes\Lambda^{n-d-4}\mathcal{R}^{\star(n-2)}
\stackrel{1\otimes\divideontimes}\longrightarrow
S^d\mathcal{R}^{\star2}\otimes\Lambda^{d+2}\mathcal{R}^{n-2}.
 \end{multline*}
The Eagon-Northcott complex is exact when the Fitting ideal $I(\varphi)$, obtained by taking all $2\times2$ determinants
of the matrix of $\varphi$, contains a regular sequence of length $(n-3)$.

For the system $\E$ that we study the map $\ell_\E$ can be split. Indeed, one easily checks that
$\ell_\E(e_u)$ and $\ell_\E(e_v)$ in sequence \eqref{264n} generate two complementary
submodules $\Lambda^2\mathcal{R}^{n-2}\subset\mathcal{R}^{2\binom{n-2}2}$. Therefore this splitting generates
two copies of the $\star$-dual Eagon-Northcott complex (in particular $\mathcal{C}_\E^\star$ is the doubling of the first
differential $\partial$) implying the following resolution of the $\star$-dual symbolic module, i.e.
dualisation of \eqref{Rcomplex} over $\mathcal{R}$ for $\E$ given by system \eqref{un}-\eqref{vn} is
 $$
0\to\mathcal{M}_\E^\star\to
\mathcal{R}^2\stackrel{\ell_\mathcal{E}}\longrightarrow\mathcal{R}^2\otimes\Lambda^2\mathcal{R}^{n-2}
\stackrel{\mathcal{C}_\E}\longrightarrow\mathcal{R}^{\star2}\otimes\mathcal{R}^2\otimes\Lambda^3\mathcal{R}^{n-2}
\stackrel{\partial^\star}\longrightarrow
S^2\mathcal{R}^{\star2}\otimes\mathcal{R}^2\otimes\Lambda^4\mathcal{R}^{n-2}\to\dots
 $$
The Fitting condition mentioned above follows from the fact that the zero set of $I(\ell_\E)$
is the tangential variety to the rational normal curve (see the initial steps of the proof of Theorem \ref{T2})
that has codimension $n-3$.

The claim that only 5-tuples of distinct indices enter the compatibility conditions can be visualised
in terms of the above complex as follows: the factor $\Lambda^3\mathcal{R}^{n-2}$
in the space of compatibility conditions refers to triples of
indices $(ijk)$ that yield the equations $E_{12ij}, E_{12jk}, E_{12ki}$ and $F_{12ij}, F_{12jk}, F_{12ki}$.
For each such 5-tuple $(12ijk)$ the number of compatibility conditions is four,
which equals the rank of the factor $\mathcal{R}^{*2}\otimes\mathcal{R}^2$.

It is not surprising that the Eagon-Northcott complex arises in our study since it is used to show that
the ideal of a rational normal curve is Cohen-Macaulay \cite[A2.19]{E}.

\section{Canonical connections for integrable PDEs}\label{App.B}

In this section we calculate  Christoffel's symbols of the canonical connections associated with 4D examples
of Section \ref{sec:ex}. This provides explicit formulae for involutive $GL(2, \bbbr)$ structures and the
associated  connections parametrised by solutions of dispersionless integrable systems
(exact solutions can be constructed by the method of hydrodynamic reductions \cite{Fer1}).

In all cases, totally geodesic $\alpha$-manifolds are projections of  integral manifolds of commuting
vector fields from the dispersionless Lax representation. These computations together with
the corresponding higher-dimensional (5D etc) counterparts (not included here) were performed in
Maple's \textsc{DifferentialGeometry} package (see arXiv:1607.01966v2).

Note that in the 4D cases considered here the normal connections coincide with totally geodesic ones; in higher
dimensions the totally geodesic connection does not exist for dKP and Adler-Shabat hierarchies (the normal connection exists),
while for the universal hierarchy it exists and coincides with the normal connection.

\subsection{Connections associated with the dKP hierarchy}

We use the notation $(x^1, x^2, x^3, x^4)=(x, y, t, z)$, note that $\Gamma_{jk}^i\neq\Gamma_{kj}^i$ in general. Not listed Christoffel symbols  are zero (unless the connection is torsion-free, in which case $\Gamma_{jk}^i=\Gamma_{kj}^i$).

\medskip
\noindent {\bf Torsion-free  $GL(2, \bbbr)$ connection} is given by
 \begin{gather*}
\Gamma^1_{13}=u_{11},\ \Gamma^1_{14}=u_{12},\ \Gamma^2_{14}=2u_{11},\ \Gamma^1_{22}=\tfrac49u_{11},\
\Gamma^1_{23}=u_{12},\ \Gamma^2_{23}=\tfrac89u_{11},\\
\Gamma^1_{24}=\tfrac{13}9u_{22}-\tfrac49u_{13},\
\Gamma^2_{24}=2u_{12},\ \Gamma^3_{24}=\tfrac43u_{11},\
\Gamma^1_{33}=\tfrac{11}9u_{13}-\tfrac29u_{22},\\
\Gamma^2_{33}=2u_{12},\ \Gamma^3_{33}=\tfrac79u_{11},\
\Gamma^1_{34}=2u_{23}-u_{14}+\tfrac43u_2u_{11},\ \Gamma^2_{34}=\tfrac{22}9u_{13}-\tfrac49u_{22},\\
\Gamma^3_{34}=3u_{12},\ \Gamma^4_{34}=\tfrac23u_{11},\
\Gamma^1_{44}=u_{33}+\tfrac{32}9u_1u_{22}-\tfrac{41}9u_1u_{13}-2u_2u_{12},\\
\Gamma^2_{44}=4u_{23}-2u_{14}+6u_2u_{11},\
\Gamma^3_{44}=\tfrac{16}3u_{13}-\tfrac73u_{22},\
\Gamma^4_{44}=4u_{12}.
 \end{gather*}

\medskip
\noindent {\bf Normal (totally geodesic) $GL(2, \bbbr)$ connection with trace-free torsion} is given by
 \begin{gather*}
\Gamma^1_{13}=u_{11},\ \Gamma^1_{14}=u_{12},\ \Gamma^2_{14}=2u_{11},\ \Gamma^1_{22}=u_{11},\
\Gamma^1_{23}=u_{12},\ \Gamma^2_{23}=2u_{11},\\
\Gamma^1_{24}=2u_{22}-u_{13},\ \Gamma^2_{24}=2u_{12},\ \Gamma^3_{24}=3u_{11},\ \Gamma^1_{31}=-\tfrac23u_{11},\
\Gamma^1_{32}=u_{12},\\
\Gamma^2_{32}=\tfrac13u_{11},\ \Gamma^1_{33}=2u_{22}-u_{13},\ \Gamma^2_{33}=2u_{12},\ \Gamma^3_{33}=\tfrac43u_{11},\
\Gamma^2_{34}=4u_{22}-2u_{13},\\
\Gamma^1_{34}=2u_{23}-u_{14}-2u_2u_{11},\
\Gamma^3_{34}=3u_{12},\ \Gamma^4_{34}=\tfrac73u_{11},\ \Gamma^1_{41}=u_{12},\
\Gamma^2_{41}=-3u_{11},\\
\Gamma^1_{42}=4u_{13}-3u_{22},\ \Gamma^2_{42}=2u_{12},\ \Gamma^3_{42}=-2u_{11},\
\Gamma^1_{43}=2u_{23}-u_{14}+3u_2u_{11},\\
\Gamma^2_{43}=3u_{13}-u_{22},\ \Gamma^3_{43}=3u_{12},\ \Gamma^4_{43}=-u_{11},\
\Gamma^1_{44}=3u_{33}-u_1u_{13}-2u_{24},\\
\Gamma^2_{44}=4u_{23}-2u_{14}+u_2u_{11},\ \Gamma^3_{44}=2u_{13}+u_{22},\ \Gamma^4_{44}=4u_{12}.
 \end{gather*}

\medskip
\noindent {\bf Totally geodesic  projective connection} is given by
 \begin{gather*}
\Gamma^1_{13}=-\tfrac12u_{11},\ \Gamma^1_{14}=-u_{12},\ \Gamma^2_{14}=-\tfrac12u_{11},\ \Gamma^1_{22}=u_{11},\
\Gamma^1_{23}=u_{12},\ \Gamma^2_{23}=\tfrac12u_{11},\\
\Gamma^1_{24}=\tfrac32u_1u_{11}+u_{22},\ \Gamma^3_{24}=\tfrac12u_{11},\
\Gamma^1_{33}=u_{22}-u_1u_{11},\ \Gamma^2_{33}=2u_{12},\\
\Gamma^1_{34}=u_{14}-2u_1u_{12}-\tfrac32u_2u_{11},\ \Gamma^2_{34}=\tfrac12u_1u_{11}+2u_{22},\ \Gamma^3_{34}=u_{12},\\ \Gamma^1_{44}=2u_{33}-u_1u_{22}-u_2u_{12}-u_{24},
\Gamma^2_{44}=2u_{14}-4u_1u_{12}-3u_2u_{11},\
\Gamma^3_{44}=3u_{13}-u_1u_{11}.
 \end{gather*}

\subsection{Connections associated with the universal hierarchy}

We again use the notation $(x^1, x^2, x^3, x^4)=(x, y, t, z)$, note that $\Gamma_{jk}^i\neq\Gamma_{kj}^i$ in general. Not listed Christoffel symbols  are zero (unless the connection is torsion-free, in which case $\Gamma_{jk}^i=\Gamma_{kj}^i$).

\medskip
\noindent {\bf Torsion-free  $GL(2, \bbbr)$ connection} is given by
 \begin{gather*}
\Gamma_{11}^2=-\tfrac23u_{11},\
\Gamma_{11}^3=-\tfrac23u_1u_{11}-u_{12},\
\Gamma_{21}^3=-\tfrac13u_{11},\\
\Gamma_{11}^4=-\tfrac13(2u_1^2+u_2)u_{11}-u_1u_{12}-u_{13},\
\Gamma_{21}^4=-\tfrac23u_1u_{11}-u_{12},\\
\Gamma_{12}^2=-\tfrac89u_{12},\
\Gamma_{22}^2=-\tfrac49u_{11},\
\Gamma_{12}^3=\tfrac19u_1u_{12}-\tfrac59u_2u_{11}-u_{13},\\
\Gamma_{22}^3=-\tfrac29u_1u_{11}-\tfrac79u_{12},\
\Gamma_{32}^3=-\tfrac29u_{11},\\
\Gamma_{12}^4=\tfrac19(u_1^2-7u_2)u_{12}-\tfrac19(7u_1u_2+3u_3)u_{11}-u_{14},\\
\Gamma_{22}^4=\tfrac19u_1u_{12}-\tfrac59u_2u_{11}-u_{13},\
\Gamma_{32}^4=-\tfrac49u_1u_{11}-\tfrac23u_{12},\\
\Gamma_{13}^3=\tfrac19(u_1^2+4u_2)u_{12}-\tfrac{10}9u_3u_{11}-u_{14},\
\Gamma_{23}^3=-\tfrac19(u_1^2+4u_2)u_{11}-u_{13},\\
\Gamma_{33}^3=-\tfrac29u_1u_{11}-\tfrac59u_{12},\
\Gamma_{43}^3=-\tfrac19u_{11},\\
\Gamma_{13}^4=\tfrac19(u_1^3+4u_1u_2)u_{12}-\tfrac19(u_1^2u_2+5u_1u_3+4u_2^2)u_{11}-u_1u_{14}-\tfrac13u_3u_{12}-u_{24},\\
\Gamma_{23}^4=\tfrac19(u_1^2+4u_2)u_{12}-\tfrac23u_3u_{11}-u_{14},\
\Gamma_{33}^4=-\tfrac19(2u_1^2+7u_2)u_{11}-\tfrac19u_1u_{12}-u_{13},\\
\Gamma_{43}^4=-\tfrac29u_1u_{11}-\tfrac13u_{12},\
\Gamma_{14}^4=\tfrac19(u_1^4+5u_1^2u_2+6u_1u_3+4u_2^2)u_{12}-u_1u_{24}\\
 \qquad -\tfrac19(u_1^3u_2+u_1^2u_3+4u_1u_2^2+12u_2u_3)u_{11}
 -(u_1^2+u_2)u_{14}-u_{34},\\
\Gamma_{24}^4=\tfrac19(u_1^3+4u_1u_2+9u_3)u_{12}-\tfrac19(u_1^2u_2-3u_1u_3+4u_2^2)u_{11}-u_1u_{14}-u_{24},\\
\Gamma_{34}^4=-\tfrac19u_1u_{22}-\tfrac89u_1u_{13}+\tfrac13u_2u_{12}-u_{23},\
\Gamma_{44}^4=-\tfrac49u_1^2u_{11}-\tfrac43u_1u_{12}-u_{22}.
 \end{gather*}

\medskip
\noindent {\bf Normal (totally geodesic) $GL(2, \bbbr)$ connection with trace-free torsion} is given by
\begin{gather*}
\Gamma_{11}^2=\Gamma_{21}^3=\Gamma_{31}^4=-u_{11},\
\Gamma_{11}^3=\Gamma_{21}^4=-u_1u_{11}-u_{12},\\
\Gamma_{11}^4=-u_1^2u_{11}-u_1u_{12}-u_2u_{11}-u_{13},\
\Gamma_{12}^1=\Gamma_{22}^2=\Gamma_{32}^3=\Gamma_{42}^4=-\tfrac13u_{11},\\
\Gamma_{12}^2=\Gamma_{22}^3=\Gamma_{32}^4=-u_{12},\
\Gamma_{12}^3=\Gamma_{22}^4=-u_2u_{11}-u_{13},\\
\Gamma_{12}^4=-u_1u_2u_{11}-u_2u_{12}-u_3u_{11}-u_{14},\\
\Gamma_{13}^1=\Gamma_{23}^2=\Gamma_{33}^3=\Gamma_{43}^4=-\tfrac13u_1u_{11}-\tfrac23u_{12},\
\Gamma_{13}^2=\Gamma_{23}^3=\Gamma_{33}^4=-u_{13},\\
\Gamma_{13}^3=\Gamma_{23}^4=-u_3u_{11}-u_{14},\
\Gamma_{13}^4=-u_1u_3u_{11}-u_1u_{14}-u_2u_{13}-u_{33},\\
\Gamma_{14}^1=\Gamma_{24}^2=\Gamma_{34}^3=\Gamma_{44}^4=
 -\tfrac13u_1^2u_{11}-\tfrac13u_1u_{12}-\tfrac23u_2u_{11}-u_{13},\\
\Gamma_{14}^2=\Gamma_{24}^3=\Gamma_{34}^4=-u_{14},\
\Gamma_{14}^3=\Gamma_{24}^4=-u_1u_{14}-u_{24},\\
\Gamma_{14}^4=-u_1^2u_{14}-u_1u_{24}-u_2u_{14}-u_{34},
 \end{gather*}
note that $\Gamma_{ij}^k=\Gamma_{i+a,j}^{k+a}$, as long as the indices are in the range.

\medskip
\noindent {\bf Totally geodesic projective connection} is given by
 \begin{gather*}
\Gamma_{12}^1=\Gamma_{13}^2=\Gamma_{14}^3=-\tfrac12u_{11},\
\Gamma_{13}^1=-\tfrac12u_1u_{11},\
\Gamma_{14}^1=-\tfrac12u_1^2u_{11}-\tfrac12u_1u_{12}-\tfrac12u_2u_{11},\\
\Gamma_{14}^2=-\tfrac12u_1u_{11}-\tfrac12u_{12},\
\Gamma_{22}^1=-u_{12},\
\Gamma_{23}^1=-\tfrac12u_2u_{11}-u_{13},\\
\Gamma_{24}^1=-\tfrac12u_1u_2u_{11}-\tfrac12u_2u_{12}-\tfrac12u_3u_{11}-u_{14},\
\Gamma_{24}^2=-\tfrac12u_2u_{11},\\
\Gamma_{24}^3=-\tfrac12u_{12},\
\Gamma_{33}^1=-u_3u_{11}-u_{14},\
\Gamma_{33}^2=-u_{13},\
\Gamma_{33}^3=u_{12},\\
\Gamma_{34}^1=-\tfrac12u_1u_3u_{11}-u_1u_{14}-u_{24}-\tfrac12u_3u_{12},\
\Gamma_{34}^2=-\tfrac12u_3u_{11}-u_{14},\\
\Gamma_{34}^4=\tfrac12u_{12},\
\Gamma_{44}^1=-u_1^2u_{14}-u_1u_{24}-u_2u_{14}-u_{34},\\
\Gamma_{44}^2=-u_1u_{14}-u_{24},\
\Gamma_{44}^3=-u_{14},\
\Gamma_{44}^4=u_{13}.
 \end{gather*}

\subsection{Connections associated with Adler-Shabat triples}

In what follows,  $i, j, k$ are pairwise distinct indices taking values $ 2, 3, 4$.

\medskip
\noindent {\bf Torsion-free $GL(2, \bbbr)$ connection} is given by (no summation unless specified):
 \begin{gather*}
\Gamma^1_{11}=\tfrac29\bigl(3R_d-R_a\bigr)-\tfrac19\sum_{i\ne 1}\sigma_{ijk}u_{ii}, \
\Gamma^1_{1i}=\tfrac13(\gamma_{jk}u_{ij}+\gamma_{kj}u_{ik})-\tfrac19R_e,\\
\Gamma^i_{11}=\tfrac19(u_i-u_j)(u_i-u_k)(\gamma_{jk}(u_{jj}-4u_{ij})+\gamma_{kj}(u_{kk}-4u_{ik})),\\
\Gamma^i_{1i}=-\tfrac29R_a-\tfrac49u_{ii}+\frac19\frac{\gamma_{jk}}{\gamma_{ji}}(u_{jj}-u_{ij})
+\frac19\frac{\gamma_{kj}}{\gamma_{ki}}(u_{kk}-u_{ik})+\tfrac59(u_{ij}+u_{ik}),\\
\Gamma^j_{1i}=\frac19\frac{\gamma_{ki}}{\gamma_{kj}}\bigl(u_{ii}-4u_{ij}+4u_{jk}-u_{kk}\bigr),\
\Gamma^1_{ii}=-\tfrac19R_f,\ \Gamma^1_{ij}=-\tfrac19R_f,\\
\!\!\!\!\! \Gamma^i_{ii}=-\tfrac13R_e+\frac{\sigma_{ijk}+2}9(\gamma_{jk}u_{jj}+\gamma_{kj}u_{kk})
+\frac29\Bigl(\gamma_{jk}-\gamma_{ij}+\gamma_{ik}-\frac{\gamma_{ij}\gamma_{ik}}{\gamma_{jk}}\Bigr)(u_{ij}-u_{ik}),\\
\Gamma^j_{ii}=-\frac{\gamma_{ij}}9\frac{\gamma_{ki}}{\gamma_{kj}}\bigl(u_{ii}-4u_{ij}+4u_{jk}-u_{kk}\bigr),\
\Gamma^k_{ij}=\frac{\gamma_{ij}}9\bigl(u_{ii}-4u_{ik}+4u_{jk}-u_{jj}\bigr),\\
\Gamma^i_{ij}=-\tfrac23\gamma_{ij}R_a-\tfrac19(\gamma_{jk}-4\gamma_{ij})(u_{jj}+u_{ik})
+\tfrac19(\gamma_{jk}+5\gamma_{ij})(u_{kk}+u_{ij}),
 \end{gather*}
where
 \begin{gather*}
\gamma_{ij}=\frac1{u_i-u_j},\
\sigma_{ijk}=\frac{\gamma_{ij}}{\gamma_{ik}}+\frac{\gamma_{ik}}{\gamma_{ij}}=
\frac{u_i-u_k}{u_i-u_j}+\frac{u_i-u_j}{u_i-u_k},\\
R_d=\sum_{i\neq1}(\gamma_{ij}+\gamma_{ik})u_{1i},\
R_e=\sum_{i\neq1}(\gamma_{ij}+\gamma_{ik})u_{ii},\
R_f=\sum_{i\neq1}\gamma_{ij}\gamma_{ik}u_{ii}.
 \end{gather*}

\medskip
\noindent {\bf Normal (totally geodesic) $GL(2, \bbbr)$ connection with trace-free torsion} is given by
 \begin{gather*}
\Gamma^1_{11}=\tfrac13(2R_b-R_a),\ \Gamma^1_{1i}=R_c,\
\Gamma^1_{i1}=-\frac13\Bigl(\frac{u_{ii}-u_{jj}}{u_i-u_j}+\frac{u_{ii}-u_{kk}}{u_i-u_k}+5R_c\Bigr),\\
\Gamma^i_{1i}=2u_{kl}-\tfrac13(R_a+R_b),\
\Gamma^i_{i1}=-u_{ii}+\frac{u_i-u_k}{u_j-u_k}u_{ij}+\frac{u_i-u_j}{u_k-u_j}u_{ik},\\
\Gamma^i_{ii}=\Gamma^1_{i1}+R_c,\ \Gamma^i_{ij}=\frac{R_b-u_{ii}-2u_{jk}}{u_i-u_j},\\
\Gamma^j_{ij}=\frac{2u_{ii}+u_{jj}-3u_{ij}-u_{ik}+u_{jk}}{3(u_i-u_j)}-\frac{u_{ii}-u_{kk}}{3(u_i-u_k)},
 \end{gather*}
where
 \begin{gather*}
R_a=u_{22}+u_{33}+u_{44},\ R_b=u_{23}+u_{24}+u_{34},\\
R_c=\frac{u_2u_{34}}{(u_2-u_3)(u_2-u_4)}+\frac{u_3u_{24}}{(u_3-u_2)(u_3-u_4)}+\frac{u_4u_{23}}{(u_4-u_2)(u_4-u_3)}.
 \end{gather*}

\medskip
\noindent {\bf Totally geodesic projective connection} is given by
 $$
\Gamma^i_{1i}=-\tfrac12u_{ii},\ \Gamma_{ij}^i=\frac{2u_{ij}-u_{ii}-u_{jj}}{2(u_i-u_j)},
 $$
recall that $\Gamma^i_{jk}=\Gamma^i_{kj}$, all other Christoffel symbols are zero.

\section*{Acknowledgements}

We thank L.\ Bogdanov, R.\ Bryant, B.\ Doubrov, M.\ Dunajski, W.\ Krynski and M.\ Pavlov for helpful discussions. We also thank the referee for useful suggestions.
BK acknowledges financial support from the LMS making this collaboration possible.

\end{document}